\documentclass{aa} 

\usepackage{graphicx} 
\usepackage{txfonts} 
\usepackage{natbib} 
 
\newcommand{\Mpc}{$h^{-1}$\thinspace Mpc}

\def\apj{ApJ} 
\def\apjl{ApJL} 
\def\apjs{ApJS} 
\def\aj{AJ} 
\def\aap{A\&A} 
\def\mnras{MNRAS}

\begin{document}    
 
\title{The biasing phenomenon}
 
\author{J. Einasto\inst{1,2,3} 
\and  L. J. Liivam\"agi\inst{1}
\and  I. Suhhonenko\inst{1} 
\and   M. Einasto\inst{1}  
}
 
\institute{Tartu Observatory, 61602 T\~oravere, Estonia 
\and  
ICRANet, Piazza della Repubblica 10, 65122 Pescara, Italy 
\and 
Estonian Academy of Sciences, 10130 Tallinn, Estonia
} 
 
\date{ Received; accepted}  
 
\authorrunning{Einasto et al.} 
 
\titlerunning{Biasing phenomenon} 
 
\offprints{J. Einasto, e-mail: jaan.einasto@to.ee} 
 
\abstract {We study biasing as a physical phenomenon by analysing 
  geometrical and clustering properties of density fields of matter
  and galaxies.}
{Our goal is to  determine the bias function using a combination of
  geometrical  and power spectrum analysis of simulated and real data.} 
{We apply an algorithm based on local densities of particles,
  $\delta$, to form simulated biased models using particles with
  $\delta \ge \delta_0$.  We calculate the bias function of model
  samples as functions of the particle density limit $\delta_0$.  We
  compare the biased models with Sloan Digital Sky Survey (SDSS)
  luminosity limited samples of galaxies using the extended
  percolation method. We find density limits $\delta_0$ of biased
  models, which correspond to luminosity limited SDSS samples.}
{Power spectra of biased model samples allow to estimate the bias
  function $b(>L)$ of galaxies of luminosity $L$.  We find the
  estimated bias parameter of $L_\ast$ galaxies,
  $b_\ast =1.85 \pm 0.15$. }
  {The absence of galaxy formation in low-density regions of the
    Universe is the dominant factor of the biasing phenomenon.
    Second largest effect is the dependence of the bias function on
    the luminosity of galaxies.  Variations in gravitational and
    physical processes during the formation and evolution of galaxies
    have the smallest influence to the bias function. }


\maketitle

\section{Introduction: biasing as a physical phenomenon}

The formation of galaxies is very complex including gravitational and
hydrodynamical processes.  On small scales hydrodynamical processes
dominate, on large scales gravitational processes.  Thus the biasing
phenomenon can be divided into local bias and large-scale bias, as
emphasised in the very detailed review by \citet{Desjacques:2018qf}.
Most papers cited by \citet{Desjacques:2018qf} were devoted to the
study of the local bias.  In this paper we are interested in the
global bias phenomenon, where gravitational processes dominate.

To quantify the large-scale galaxy bias various statistical tools were
applied as the galaxy autocorrelation function, starting from
\citet{Kaiser:1984}, \citet{Bardeen:1986} and
\citet{Szalay:1988aa}. Another statistic used to study the
relationship between matter and galaxies is the void probability
function (VPF), as done by \citet{Gramann:1990tg},
\citet{Einasto:1991fq} and more recently by \citet{Walsh:2019aa}.
Presently the power spectrum analysis has been preferred.  The power
spectrum analysis has several  advantages over the
correlation function analysis \citep{Feldman:1994fr}.  The power
spectrum measures the fractional density contributions on different
scales, and is a natural quantity to describe the density field,
especially on large scales.

Different authors used different data and different methods to the
estimation of power spectra, which led to various definitions of the
bias parameter.  Geometrical properties of the distribution of
galaxies and matter were discussed separately using a large variety of
methods.  This area of research is very rich, as seen from discussions
on the recent symposium of cosmic web \citep{VandeWeygaert:2016zt}.
The connection of determinations of power spectra with geometrical
properties of the cosmic web were discussed rarely.

In the present paper we try to get a more general view of the biasing
phenomenon in the context of structure of the cosmic web.
Biasing phenomenon is a physical problem of the relation between
distributions of matter and galaxies on large scales.  In discussing
the biasing phenomenon we assume that gravity is the dominating force
which determines the formation and evolution of the cosmic web on
large scales.  According to the presently accepted $\Lambda$ cold dark
matter ($\Lambda$CDM) model the primordial density field forms a
statistically homogeneous, isotropic and almost-Gaussian random field.
Density waves of different scales began with random and uncorrelated
spatial phases.  As the density waves evolve, they interact with
others in a non-linear way. This interaction leads to the generation
of non-random and correlated phases which form the spatial pattern of
the present cosmic web with clusters, filaments, sheets and voids.
Matter flows out from under-dense regions towards over-dense regions,
which changes the pattern of the evolving cosmic web. In high-density
regions there exists conditions favourable for the formation of
galaxies.

We consider the bias function $b$ as a fundamental cosmological
function, which relates quantitatively differences between
distributions of matter and galaxies.  The numerical value of the
bias function can be found by the power spectrum analysis, the
relation between galaxies and matter can be found using geometrical
properties of the cosmic web.  We discuss
first shortly basic physical processes involved in the biasing phenomenon:
the formation of galaxies in the cosmic web, the phase synchronisation
of density perturbations, and the evolution of voids.

\subsection{Formation of the biasing concept}

An important element of the classical version of the cosmology
paradigm is the distribution of galaxies.  Available data on the
distribution of galaxies on sky suggested that this distribution is
essentially a random one (field galaxies) to which some clusters and
perhaps even superclusters were added, see the angular distribution of
the numbers of galaxies brighter than $B \approx 19$ by
\citet{Seldner:1977aa}.  The angular distribution of galaxies can be
considered as a random Gaussian process, and described by the angular
correlation function of galaxies, as done by 
\citet{Peebles:1973a}, \citet{Peebles:1974a}, \citet{Peebles:1975} and
\citet{Davis:1983yq}.

In 1970s the number of galaxies with measured redshifts was
sufficiently large to study the distribution of galaxies in three
dimensions.  The topic was discussed in the IAU Symposium 79 ``Large
Scale Structure of the Universe'' in Tallinn by \citet{Joeveer:1978pb},
\cite{Tarenghi:1978}, \cite{Tifft:1978} and \citet{Tully:1978}.
Three-dimensional data demonstrated that the distribution of galaxies
and clusters of galaxies is filamentary and that there are almost no
galaxies outside filaments.  A theory of the formation of galaxies due
to gravitational instability was suggested by \citet{Zeldovich:1970}.
Numerical simulations in the framework of this model by
\citet{Doroshkevich:1980} demonstrated the formation of a cellular
network of high- and low-density regions.  Due to the similarity of
the observed large-scale distribution of galaxies to the structure,
found in simulations, the structure was called ``cellular''
\citep{Joeveer:1977, Joeveer:1978dz}.  Subsequently more general terms
``supercluster-void network'' \citep{Einasto:1980} and ``cosmic web''
\citep{Bond:1996fv} were used.

\citet{Joeveer:1977, Joeveer:1978dz} estimated that knots, filaments
and sheets fill only about one per cent of the whole volume of the
universe, the rest forms voids.  Authors noticed that gravity works
very slowly and it is very unlikely to evacuate completely such large
volumes as cell interiors: there must exist unclustered matter in
voids.  In this way the difference between distributions of matter and
galaxies was detected.  The structure of superclusters and voids was
investigated quantitatively by \citet{Zeldovich:1982kl} by comparing
distributions of simulated particles with real galaxies.  The
multiplicity test confirmed the existence of a smooth population of
void particles in simulations.  The multiplicity test also showed the
absence of a large low-density population of void galaxies in the
observed sample.  To explain these differences
\citet{Zeldovich:1982kl} assumed that the matter density in voids and
sheets is too small to start galaxy formation.  The term ``biasing''
was suggested later by \citet{Kaiser:1984} to note the difference
between correlation functions of clusters of galaxies in respect to
galaxies.  Subsequently this term was used in a more general case to
denote differences in the distribution of galaxies and matter
\citep{Davis:1985}.

\subsection{Formation of galaxies in the web}

Galaxy formation is a two-stage process --- gravitating material in
the Universe condenses first into DM  halos
\citep{White:1978}.  To form a galaxy, the density of matter must
exceed a critical value, the Press-Schechter limit
\citep{Press:1974fk}.  This result is confirmed by hydrodynamical
models of galaxy formation \citep[for early model see ][]{Cen:1992kx}.  The
arguments by White and Rees are supported by direct observational
evidence --- all galaxies are DM  dominated, especially dwarf
galaxies \citep{McConnachie:2012aa}.  The luminous content of galaxies
results from the combined action of gravitational and hydrodynamical
processes within potential wells provided by the DM .

Arguments by Zeldovich, White and Rees lead to a simple biasing model,
where galaxies do not form in low-density regions at all, or are too
faint, to be included into flux-limited galaxy surveys.

\subsection{Phase synchronisation}

The expansion of the Universe in its early phase is an adiabatic
process \citep{Zeldovich:1970, Peebles:1982}. The growth of adiabatic
perturbations proceeds at a low temperature of the primordial ``gas'',
and the flow of particles is very smooth \citep{Zeldovich:1970,
  Zeldovich:1978, Zeldovich:1982kl}.  Smooth initial perturbations
develop into the non-linear stage and dense regions will be built up
by the concentration of matter into caustics by intersection of
particle trajectories \citep{Zeldovich:1978, Zeldovich:1982kl,
  Arnold:1982aa}.  In this way the skeleton of the cellular cosmic web
is formed.  First three-dimensional numerical simulation by
\citet{Doroshkevich:1982fk} showed only the formation of very large
systems with bulky connections without a web of fine filaments.  This
simulation was made under the assumption that DM  is made of
massive neutrinos, this is the hot dark matter (HDM) model.  Weakly
interacting massive particles, called cold dark matter (CDM), were
suggested by \citet{Peebles:1982cr}.  Quantitative analysis of a CDM
model by \citet{Melott:1983} showed that the CDM model is in good
agreement with observations.  In particular, all quantitative tests,
applied to the HDM model by \citet{Zeldovich:1982kl}, showed that the
CDM model is in good agreement with observed samples of galaxies.  In
the CDM model the intersection of particle trajectories leads directly
to the early formation of thin filaments and knots, as shown by
\citet{Melott:1983}, and subsequently studied in more detail by
\citet{White:1987}, \citet{Kofman:1990aa} and \citet{Bond:1996fv}.

When the presence of voids was discovered, \citet{Dekel:1986kx}
assumed that voids can be populated with dwarf galaxies.  However,
observations suggested that voids are marked by the absence of both
normal and dwarf galaxies \citep{Einasto:1989cr, Lindner:1995ui,
  Lindner:1996tu, Peebles:2001kl,  Tinker:2006fr}. 

The absence of dwarf galaxies in voids has a simple explanation.  The
growth of density perturbations is an acoustic phenomenon and can be
studied by the wavelet technique \citep{Einasto:2011dq,
  Einasto:2011kx}. Voids are regions in space where due to phase
synchronisation medium- and large-scale density waves combine in
similar under-density phases.  Here the growth of {\em all}
small-scale density perturbations, responsible for the formation of
galaxies, is suppressed.  Small-scale
density perturbations form initially everywhere, but in regions of
under-dense phases of large and medium perturbations the density
contrast of small-scale perturbations decreases during the evolution.
Galaxies, clusters and superclusters form in regions where medium- and
large-scale density waves combine in similar over-density phases.
Near maxima of large-scale density perturbations medium and
small-scale perturbations grow.  This leads to the formation of
numerous halos and subhalos around the high- and medium-density peaks.
It is possible that just the phase synchronisation leads to the
formation of galaxy systems with centrally located giant galaxies
surrounded by dwarf galaxies.  The formation of satellite galaxies and
relations between satellite and main galaxies is presently a subject
of intensive studies, both observational and theoretical; for a recent
review see \citet{Wechsler:2018fj}.

\subsection{Evolution of voids}

Gravity works slowly and there exist always particles which are
more-or-less smoothly distributed in low-density regions.
\citet{Sheth:2004ly}, \citet{Rieder:2013aa} and 
\citet{Aragon-Calvo:2016aa} among  others 
investigated how galaxies form and evolve inside the cosmic web.
Galaxies accrete star-forming gas at early times via the network of
primordial filaments.  The flow of gas along filaments continues.  It
is evident, that not all matter in filaments (and walls) is presently
located in halos --- the process is still going on.  Density
distributions of SDSS samples can be compared with density
distributions models.  \citet{Cautun:2014qy}, \citet{Falck:2015pd} and
\citet{Ganeshaiah-Veena:2019fk} highlighted regions belonging to
simulated voids, walls, filaments and halos.  In all models there
exists large under-dense regions, where halo (and thus galaxy)
formation is not possible.  These studies suggest that the fraction of
particles in low-density regions, not associated with halos, is about
25 - 30~\% of all particles. This non-clustered matter forms a
more-or-less uniformly distributed medium, part of it is located in
weak filaments of  dark and baryonic matter \citep{Aragon-Calvo:2010ve}.

\subsection{Goal of the present paper}

The goal of this paper is study  biasing as a physical
phenomenon and to estimate the bias function using clustering and
geometrical properties of the distribution of DM and galaxies.  We
divide this task to three subtasks: (i) the generation of biased model
distributions of matter and the study of geometrical properties of the
distribution of DM, simulated and real galaxies; (ii) the calculation
of power spectra of simulated biased models and the determination of
the bias function using simulated biased models; (iii) the comparison
of geometrical properties of simulated and real galaxy samples to find
the simulated model which best represents the observed
distribution. In simulations full data on the distribution of
particles, selection and boundary effects etc.  are known, and the
comparison of biased and full models is fully differential. As a
result of the comparison we find the relation between the biased model
selection parameter and galaxy sample selection parameter. The bias
function is a function of the selection parameter, used to define
biased models to simulate galaxies.  This method circumvents the main
difficulty of the power spectrum analysis of the biasing phenomenon:
from observations we can determine the power spectrum directly only
for galaxies of various luminosity, but not for matter.  Following
this idea we have to decide: (i) how to construct biased model
samples, and (ii) how to compare biased models with observations.

We shall use a simple biased DM  simulation model and divide
matter into a low-density population with no galaxy formation or
populated with galaxies below a certain luminosity limit, and to a
high-density population with clustered matter, associated with
galaxies above the luminosity limit.  From observations we get
information on the distribution of galaxies at the present epoch
(actually the mean age of our observational SDSS sample corresponds to
age at redshift $z=0.1$). Following this consideration we use
present-day (Eulerian) particle local densities, $\delta$, and label
each particle with this density value.  Halos surrounding
galaxies like our own Galaxy and M31 have an effective radius of the
order of 1~\Mpc\ \citep{Einasto:1974fv, Karachentsev:2002b}.  Inside
such halos hydrodynamical processes leading to the formation of
visible galaxies are dominant.  In our study we shall use density fields
with the size of individual cells 1~\Mpc, thus all the details of
galaxy formation are hidden inside  cells.  Local density is
expressed in mean density units of the sample. It is a dimensionless
quantity and independent on particle mass and galaxy luminosity.

We apply a sharp particle density limit, $\delta_0$.  Biased model
samples include particles with density labels, $\delta \ge \delta_0$.
These samples are found from the full DM  sample by exclusion
particles of density labels less than the limit $\delta_0$.  In this
way biased model samples mimic observed samples of galaxies, where
there are no galaxies fainter  than a certain luminosity limit.  We
use a series of particle density limits $\delta_0$ to find limits,
which correspond in the best way to observational samples of galaxies.

To investigate geometrical properties of the cosmic web and to compare
models with observations we apply the extended percolation analysis,
developed by \citet{Einasto:1986aa}, \citet{Einasto:1987kw} and
\citet{Einasto:2018aa}.  The extended percolation analysis is aimed to
describe geometrical properties of the whole cosmic web. It is a
complementary to methods which aims are descriptions of properties of
elements of the cosmic web, such as knots, filaments, sheets and
voids.  The extended percolation method allows the comparison of
samples with very different border configurations, such as
observational samples with conical shell borders and cubic model
samples.  The extended percolation analysis uses for comparison
completely different properties of the cosmic web than the power
spectrum analysis, and is suitable to find proper biased models for
comparison with observational data.

As a basic reference model sample we use a numerical simulations of
the evolution of the web applying $\Lambda$CDM cosmology in a box of
size 512~\Mpc, almost equal in volume to the volume of the
flux-limited SDSS main galaxy survey, (509~\Mpc)$^3$
\citep{Liivamagi:2012aa}.  We use cosmological parameters: Hubble
parameter $H_0 = 100 h$~km~s$^{-1}$~Mpc$^{-1}$, matter density
parameter $\Omega_{\mathrm{m}} = 0.28$, and dark energy density
parameter $\Omega_{\Lambda} = 0.72$.  For comparison we also use halo
mass selected samples from the Horizon Run 4 (HR4) simulation by
\citet{Kim:2015aa} in a box of size $512$~\Mpc.  As observational
data we shall use absolute magnitude (volume) limited SDSS samples
with limits $M_r - 5\log h =-18.0,~-19.0,~-20.0,~-21.0$.

The paper is organised as follows.  In the next Section we describe
the calculation of the density fields of observed and simulated
samples, and the method to find clusters, voids and their parameters.
In Section 3 we investigate geometrical properties of the cosmic web
as delineated by matter and galaxies using the extended percolation
method.  In Section 4 we calculate power spectra and estimate the bias
function of model samples.  In Section 5 we compare observed and
simulated clusters and voids, and find biased models which correspond
to luminosity selected SDSS galaxies in the best way.  In Section 6 we
discuss our results.  The last Section brings our main conclusions.

\section{Data and methods}

\subsection{Particle density selected   model samples}

Simulations of the evolution of the cosmic web were performed in a box
of size $L_0=512$~\Mpc, with resolution $N_{\mathrm{grid}} = 512$ and
with $N_{\mathrm{part}} = N_{\mathrm{grid}}^3$ particles. The initial
density fluctuation spectrum was generated using the COSMICS code by
\citet{Bertschinger:1995}, assuming cosmological parameters
$\Omega_{\mathrm{m}} = 0.28$, $\Omega_{\Lambda} = 0.72$,
$\sigma_8 = 0.84$, and the dimensionless Hubble constant $h = 0.73$.
To generate initial data we used the baryonic matter density
$\Omega_{\mathrm{b}}= 0.044$ (\citet{Tegmark:2004}).  Calculations
were performed with the GADGET-2 code by \citet{Springel:2005}.

The accepted $\sigma_8$ for the model is in good agreement with 
 $\sigma_8$ determinations for matter, see  
\citet{Planck-Collaboration:2018kx} for Planck 2018 results. Spectra
of CMB temperature and polarisation in combination with CMB
gravitational lensing yield $\sigma_8 = 0.8111 \pm 0.0060$. If data on
BAO at lower redshifts is added, then the result is
$\sigma_8 = 0.8102 \pm 0.0060$.  \citet{Zubeldia:2019aa} derived
cosmological constraints using Planck sample of clusters, detected via
the Sunyaev-Zeldovich (SZ) effect.  Authors find
$\sigma_8 = 0.76 \pm 0.04$.  The signal from CMB comes from redshift
$z=1100$, SZ clusters have characteristic redshift $z \approx 0.2$.
Thus data from different distances are in very good agreement.

{\scriptsize 
\begin{table}[ht] 
\caption{L512 particle density limited models.} 
\label{Tab2}                         
\centering
\begin{tabular}{lrlll}
\hline  \hline
Sample   & $\delta_0$& $F_C$&$FF_C$&$b(\delta_0)$  \\  
\hline  
(1)&(2)&(3)&(4)&(5)\\ 
\hline  
L512.00   &  0.0 & 1.000 &1.0000& 1.000  \\
L512.05 &  0.5 & 0.901&0.5163&  1.199   \\
L512.1   &  1.0 & 0.797 &0.3434&  1.302 \\
L512.2   &  2.0 &  0.678 &0.2159& 1.429 \\
L512.5   &  5.0 &  0.516 &0.10743& 1.635 \\
L512.7   &  7.5 &  0.449 &0.07665&  1.740  \\
L512.10 &  10.0 & 0.4036 &0.05972& 1.820  \\
L512.15 &  15.0 & 0.3435 &0.04138& 1.943  \\
L512.20 &  20.0 & 0.3011 &0.03146& 2.039 \\
L512.50 &  50.0 & 0.1831 &0.01169& 2.432  \\
L512.100&100.0& 0.1046 &0.00467& 2.970\\
\hline 
\end{tabular} 
\tablefoot{
The columns are:
(1): sample name; 
(2)  particle density limit $\delta_0$;
(3): fraction of numbers of particles in the sample,  $F_{C}$, equal to the
number density of clustered particles per cubic \Mpc;  
(4): total filling factor of all clusters at density threshold
$D_t=0.1$, $FF_C$;
(5): bias parameter, calculated from power spectra of biased models
with particle density limits  $\delta_0$.}
\end{table} 
} 

We calculated for all simulation particles and all simulation epochs
local density values at particle locations, $\delta$, using positions
of 27 nearby particles.  Densities were expressed in units of the mean
density of the whole simulation.  In this paper we used particle
density selected samples at the present epoch. Biased model samples
contain particles above a certain limit, $\delta \ge \delta_0$, in
units of the mean density of the simulation.  For the analysis we used
density limits
$\delta_0=0.5,~1.0,~2.0,~5.0,~7.5,~10.0,~15.0,~20.0,~50.0,~100.0$.
Particle density selected samples are called biased model samples and
are denoted as L512.$i$, where $i$ denotes the particle density limit
$\delta_0$.  The full DM  model includes all particles and
corresponds to particle density limit $\delta_0 = 0$, thus it is
denoted as L512.00.  Main data on biased model samples are given in
Table~\ref{Tab2}.  We give in the Table also the fraction of
particles, $F_C=N_C/N_{\mathrm{part}}$, where $N_C$ is the number of
particles with density limit $\delta \ge \delta_0$, and
$N_{\mathrm{part}}$ is the total number of particles in simulation.
$F_C$ is equal to the number density of selected particles,
$\delta \ge \delta_0$, per cubic \Mpc, $Dens$.  We give also the total
filling factor of over-density regions at density threshold $D_t=0.1$,
$FF_C$, and the bias parameters $b(\delta_0)$,  calculated
using Eq.~(\ref{gert}) below.

The use of the local density as the only parameter to determine the
fate of particles in the web is a simplification, see the distribution
of particles in regions of different density by \citet{Cautun:2014qy}
and \citet{Ganeshaiah-Veena:2019fk}.  However, as shown among others
by \citet{Tinker:2009bh}, just the local density, not the global one,
is essential in the determination of the formation of galaxies inside
DM  haloes.  In observational SDSS samples and comparison HR4
model samples we used sharp limits of absolute magnitudes and halo
masses.  The formation of galaxies inside DM halos is determined by a
variety of processes. For this reason particles of slightly various
densities can be located in halos of fixed lower mass limit and the
actual particle density limit is fuzzy. To take this effect into
account we made additional calculations with L512 models with fuzzy
particle density limits, see below.

\subsection{Luminosity limited SDSS galaxy samples}

We use luminosity limited (usually called volume-limited) galaxy
samples by \citet{Tempel:2014uq}, selected from the data release DR10
of the SDSS galaxy redshift survey \citep{Ahn:2014aa}. Data on four
luminosity limited SDSS samples are given in Table~\ref{Tab1}.
Limiting absolute magnitudes in red $r-$band, $M_r$, maximum comoving
distances, $d_{\mathrm{lim}}$, and numbers of galaxies in samples,
$N_{\mathrm{gal}}$, are taken from \citet{Tempel:2014uq}; volumes of
samples, $V_0$, and sample lengths, $L_0$, are determined by counting
cells inside the conical survey volume, and the maximum lengths of
over-density regions (clusters in our terminology).  SDSS samples with
$M_r - 5\log h $ luminosity limits $-18.0$, $-19.0$, $-20.0$, and $-21.0$ are
called SDSS.18, SDSS.19, SDSS.20 and SDSS.21.  Respective luminosity
limits in Solar units were calculated using the absolute magnitude of
the Sun in $r-$band, $M_\odot= 4.64$ \citep{Blanton:2007aa}.  We give
in the Table also the number density of sample galaxies per cubic
\Mpc, $Dens$. 
SDSS galaxy samples are conical and have different sizes: the volume
of the sample SDSS.21 is 52 times larger than the volume of the sample SDSS.18.

{\scriptsize 
\begin{table}[ht] 
\caption{SDSS luminosity limited samples.} 
\label{Tab1}                         
{\centering
\begin{tabular}{lccccrl}
\hline  \hline
Sample   &$M_r$  & $d_{\mathrm{lim}}$ & $L_0$& $V_0$ &
                                                  $N_{\mathrm{gal}}$ &$Dens$ \\  
\hline  
(1)&(2)&(3)&(4)&(5)&(6)&(7)\\ 
\hline  
SDSS.18 & $-18.0$ & 135 &  243 &  $117^3$ & 49\,860 & 0.0311\\
SDSS.19 & $-19.0$ & 211 & 379 & $188^3$  & 105\,041 &   0.0158\\
SDSS.20 & $-20.0$ & 323 & 581 & $290^3$ & 163\,094 &  0.00669\\
  SDSS.21 & $-21.0$ & 486 & 865 &$438^3$  & 125\,016 & 0.00149\\
  
\hline 
\end{tabular} 
}
 \tablefoot{
 The columns are:
(1): sample name; 
(2): absolute $r-$magnitude limit, $M_r - 5\log h $;
(3): maximum comoving distance $d_{\mathrm{lim}}$ in \Mpc;
(4): effective length of the sample in \Mpc;
(5): volume of the sample in  (\Mpc)$^3$;
(6): number of galaxies in a sample;
(7): number density of galaxies per cubic \Mpc.
}
\end{table} 
}

\subsection{Halo mass limited model samples}

To check the comparison of observed and biased model data we applied
the extended percolation analysis also for a series of halo mass
limited model samples, taken from the Horizon Run 4 simulation by
\citet{Kim:2015aa}.  This simulation was made in a cubic box of size
$3150$~\Mpc, using $6300^3$ particles, in a $\Lambda$CDM cosmology
with $\Omega_m = 0.26$, $\Omega_b=0.044$, $\Omega_\Lambda = 0.74$,
amplitude parameter $\sigma_8 = 0.794$, and current Hubble
expansion constant $H_0 = 100 h$~km/s/Mpc, where $h=0.72$.  We
selected from this simulation halos for the present epoch in a box of
size $L_0 = 512$~\Mpc.  Halos were found containing at least 30
particles, minimal mass of halos is
$M_s = 2.706\times 10^{11}$~$h^{-1} M_\odot$.  This simulation
contains 8 particles per cell of length 1~\Mpc, mean density of matter
per cell is $M_{mean} = 0.7216\times 10^{11}$~$h^{-1} M_\odot$.  We
use four halo mass limited samples from HR4 simulation. Main data on
HR4 samples are given in Table~\ref{Tab3}.  We give in the Table also
the fraction of mass in the clustered population, $F_C$, and the number
density of halos per cubic \Mpc, $Dens$.

{\scriptsize 
\begin{table}[ht] 
\caption{HR4 halo mass limited samples.} 
\label{Tab3}                         
\centering
\begin{tabular}{lcrcc}
\hline  \hline
Sample   & $M_h$ & $N_{\mathrm{halo}}$ & $F_C$& $Dens$   \\  
\hline  
(1)&(2)&(3)&(4)&(5)\\ 
\hline  \\
HR4.11 & $2.71\times 10^{11}$ & 1\,561\,724 &0.3729& 0.01164  \\
HR4.12 & $1.00\times 10^{12}$ & 254\,067 & 0.3184&0.00189   \\
HR4.123 & $3.00\times 10^{12}$ & 223\,653 &0.2659&  0.00166  \\
HR4.13 & $1.00\times 10^{13}$ & 53\,484 &0.2011& 0.00040  \\
\hline 
\end{tabular} 
\tablefoot{
The columns are:
(1): sample name; 
(2): halo mass limit in $h^{-1}~M_\odot$;
(3): number of halos in  sample;
(4): fraction of clustered mass in the sample,  $F_{C}$;
(5): Number density of halos per cubic \Mpc.
}
\end{table} 
}

\subsection{Calculation of percolation functions} 
 
The first step in the extended percolation analysis is the calculation
of density fields.  Here we applied the $B_3$ spline
\citep[see][]{Martinez:2002fu}.  This function is different from zero
only in the interval $x\in[-2,2]$.  To calculate the high-resolution
density field we use the kernel of the scale, equal to the cell size
of the simulation, $L_0/N_{\mathrm{grid}}=1$~\Mpc, where $L_0$ is the
size of the simulation box, and $N_{\mathrm{grid}}$ is the number of
grid elements in one coordinate.  The smoothing with index $i$ has a
smoothing radius $r_i= L_0/N_{\mathrm{grid}} \times 2^i$. The
effective scale of smoothing is equal to $ r_i$.  We applied this
smoothing up to index 6.  For our L512 model smoothing indexes $i=1$,
2 and 3 correspond to $B_3$ kernels of radii $R_B =2$, 4 and 8~\Mpc,
respectively. The $B_3$ kernel of radius $R_B=1$~\Mpc\ corresponds to
a Gaussian kernel with dispersion $R_G = 0.6$~\Mpc\
\citep{Tempel:2014uq}.  Non-smoothed density field corresponds to
kernel $R_B =1$~\Mpc.   Densities were expressed in mean density
units. 

The calculation of percolation functions consists of several 
steps.  We scanned  density fields in a range of threshold
densities from $D_t=0.1$ to $D_t=10$ in mean density units to find
over- and under-density systems, called  clusters and voids,
respectively.  We used a logarithmic step of densities,
$\Delta \log D_t = 0.02$.  Two cells of the same type are considered
as members of a system if they have a common sidewall.  For each cluster and void
we calculate the centre coordinates, $x_c, y_c, z_c$ (mean values of
extreme $x, y, z$ coordinates); sizes along coordinate axes,
$\Delta x,~ \Delta y,~ \Delta z$ (differences between extreme
$x, y, z$ coordinates); geometrical diameters,
$L_{\mathrm{geom}} = \sqrt{(\Delta x)^2 + (\Delta y)^2 + (\Delta
    z)^2}$; maximal sizes along coordinate axes,
$L_{\mathrm{max}} = \max(\Delta x,\Delta y,\Delta z)$; volumes, $V_C$,
defined as the volume of space where the density is equal or greater
than the threshold density $D_t$; total masses (or luminosities),
$M_t$, i.e. the masses (luminosities) inside the density contour $D_t$
of the cluster, both in mean density units.

During the cluster search we found the cluster with the largest volume
for the given threshold density, and stored for each threshold density
the number of clusters found, and data on the largest cluster: the
geometrical diameter (the maximal size along coordinate axes), the
volume, and the total mass (luminosity).  Diameters (lengths) of
largest clusters, ${L}(D_t) = L_{\mathrm{max}}$, filling factors of
largest clusters, $\mathcal{F}(D_t) = V_{\mathrm{max}}/V_0$, and
numbers of clusters at the threshold density, $\mathcal{N}(D_t)$, as
functions of the threshold density, $D_t$, were used as percolation
functions.  Diameters are expressed in \Mpc, volumes (actually the
filling factors) are expressed in units of the volume of the whole
sample, $V_0$.  During the search of high- and low-density systems we
excluded very small systems, using exclusion volume limit,
$N_{\mathrm{lim}}=50$ or 500 computation cells (cubic \Mpc).  Length
functions ${L}(D_t)$ and filling factor functions $\mathcal{F}(D_t)$
are not influenced by the choice of $N_{\mathrm{lim}}$.

We use as a percolation function the volume of the largest cluster,
$V_{\mathrm{max}}$, not its mass, $M_{\mathrm{max}}$.  The mass
function depends on the mass concentration inside halos.  The volume is
free from this dependence, here the question is whether the cluster
lies above or below the limit dividing particles or halos to over- and
under-dense regions.  This aspect is treated by the fuzziness of the
particle selection limit.

A similar procedure was applied to find the largest low-density
regions or voids in our terminology.

\section{Geometrical properties of density fields of matter and galaxies} 

In this Section we compare distributions of matter and galaxies using
the extended percolation method.  We pay special attention to
similarities and differences between distributions of DM and galaxies,
both simulated and real.

\begin{figure*}
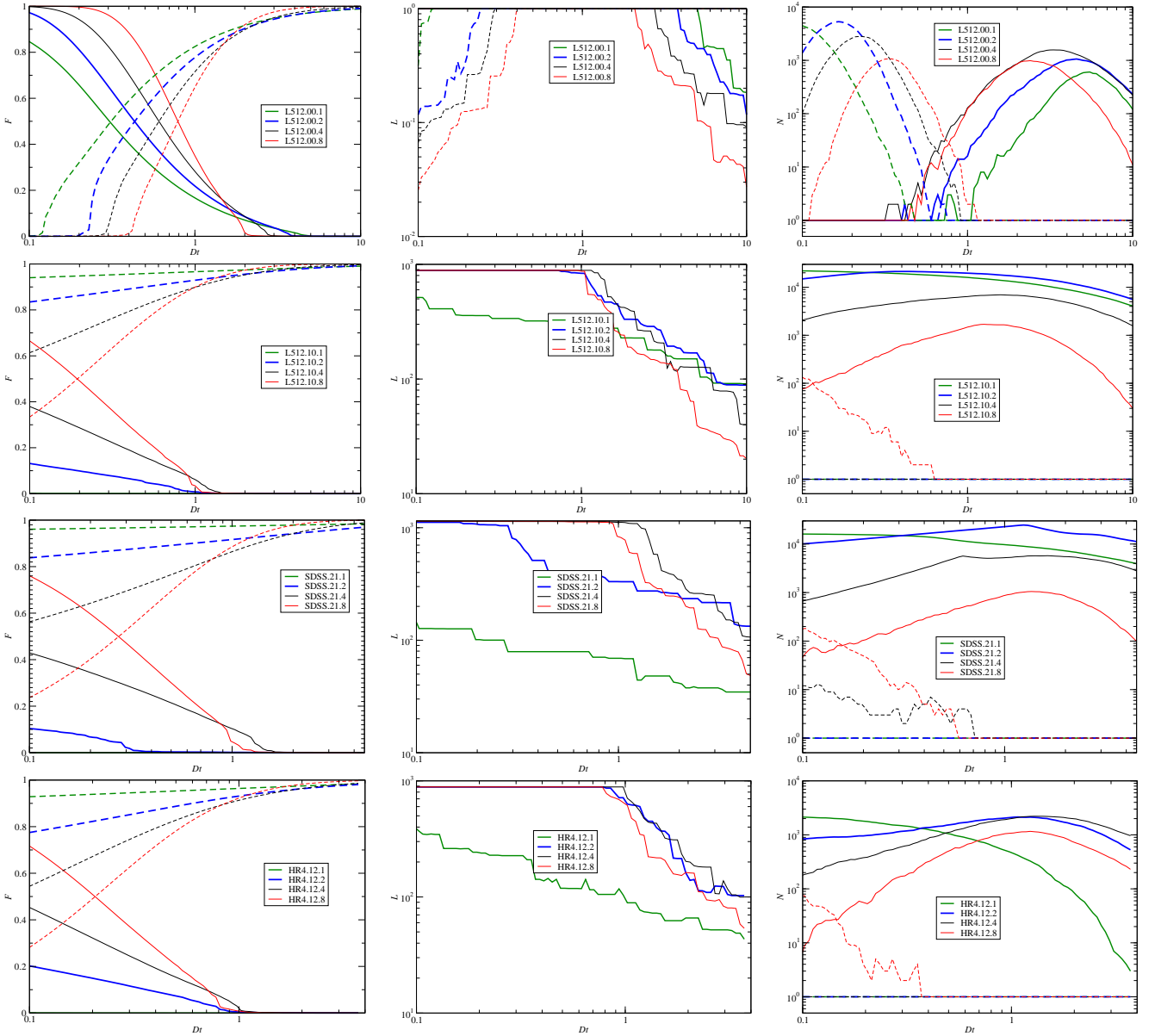
  
\centering 
\hspace{2mm}  
 \resizebox{0.30\textwidth}{!}{\includegraphics*{L512Bias0_Vmax-D0_4.eps}}
\hspace{2mm} 
 \resizebox{0.30\textwidth}{!}{\includegraphics*{L512Bias0_Lmax-D0_4.eps}}
\hspace{2mm} 
\resizebox{0.30\textwidth}{!}{\includegraphics*{L512Bias0_N-D0_4.eps}}\\
 \hspace{2mm}
\resizebox{0.30\textwidth}{!}{\includegraphics*{L512Bias10_Vmax-D0_4.eps}}
 \hspace{2mm} 
\resizebox{0.30\textwidth}{!}{\includegraphics*{L512Bias10_Lmax-D0_new.eps}}
\hspace{2mm} 
\resizebox{0.30\textwidth}{!}{\includegraphics*{L512Bias10_N-D0_new.eps}}\\
\hspace{2mm}  
\resizebox{0.30\textwidth}{!}{\includegraphics*{dr10_m21_Vmax-D0_bias45.eps}}
 \hspace{2mm}  
  \resizebox{0.30\textwidth}{!}{\includegraphics*{dr10_m21_Lmax-D0_bias45.eps}}
\hspace{2mm}  
  \resizebox{0.30\textwidth}{!}{\includegraphics*{dr10_m21_N-D0_bias45.eps}}\\
  \hspace{2mm}
\resizebox{0.30\textwidth}{!}{\includegraphics*{HR4.12_Vmax-D0_new.eps}}
 \hspace{2mm}
\resizebox{0.30\textwidth}{!}{\includegraphics*{HR4.12_Lmax-D0_new.eps}}
 \hspace{2mm}
\resizebox{0.30\textwidth}{!}{\includegraphics*{HR4.12_N-D0_new.eps}}\\
  \hspace{2mm}
  \caption{Percolation functions for unbiased L512.00 samples, biased
    L512.10 samples, SDSS.21 and HR4.12 samples, from top to bottom
    panels, respectively.  {\em Left panels} are for filling factor
    functions, {\em middle panels} for cluster length functions, {\em
      right panels} for number functions.  Functions for clusters are
    plotted with solid lines, and for voids with dashed lines.}
\label{fig:Fig14} 
\end{figure*}

\subsection{Percolation analysis of distributions of matter and galaxies}

The original percolation analysis was designed to measure the
connectivity of percolating clusters \citep{Stauffer:1979aa}, and was
used by \citet{Zeldovich:1982kl} and \citet{Melott:1983} to
investigate the connectivity of HDM and CDM models, respectively.  The
extended percolation method was designed by \citet{Einasto:1986aa},
\citet{Einasto:1987kw} and \citet{Einasto:2018aa} to compare more
general geometrical properties of models with observations. The
density field is divided into high- and low-density regions, using a
variable density threshold.  Each element of the cosmic web belongs to
a high- or low-density region, depending on the threshold.  In this
way the extended percolation analysis is a method to study various
geometrical properties of the whole cosmic web over a large range of
densities, and to compare models with observations.

Percolation functions describe how geometrical properties, such as
sizes and volumes of largest clusters and voids, depend on the
threshold used to divide the density field into high- and low-density
regions.  At high threshold density only the highest peaks (central
regions of the largest cluster) are considered as over-density
regions, and sizes and filling factors of largest clusters are small.
As the density threshold decreases outer lower density regions of the
clusters are included as parts of clusters, and  lengths and filling
factors of  clusters increase.  At certain density threshold the cluster
merges with a neighbouring cluster, and the length and filling factor
of the largest cluster increase.  After several mergers the largest
cluster spans the whole sample, i.e. it percolates  \citep{Liivamagi:2012aa}.

Fig.~\ref{fig:Fig14}  presents percolation functions: filling
factors of largest clusters and voids, $\mathcal{F}(D_t)$, lengths of
largest clusters and voids, ${L}(D_t)$, and numbers of clusters and
voids, $\mathcal{N}(D_t)$, as functions of
the threshold density, $D_t$, to divide the density field into over-
and under-density regions.  Functions are given for the full unbiased
model L512.00, and for simulated and real galaxy samples, represented
by the biased model L512.10, by the the luminosity limited sample
SDSS.21, and by the halo mass limited model HR4.12. 
These simulated and real galaxy samples correspond approximately to
$L_\star$ galaxies, see below.

The basic difference between the full model sample L512.00 and galaxy
samples lies in the presence of the weak filamentary web in the
L512.00 sample, which is absent in galaxy samples.  Percolation
functions describe this difference in various ways.  In the full model
L512.00 weak filaments and sheets fill the whole space, as seen in
Fig.~\ref{fig:Fig11} below, which isolate small low-density regions
(voids).  This property of the web is characterised in the number
function as follows: at small threshold densities the number of voids
of the model L512.00 is high, see the upper right panel of
Fig.~\ref{fig:Fig14}.  In contrast, at small threshold density
high-density regions form one connected system --- cluster, since
low-density filaments join knots of the density field into one
connected system.  With increasing threshold density some filaments
became fainter than the threshold density, the connected cluster
splits to smaller units. Simultaneously there appear tunnels between
small voids and voids merge. This leads to a rapid increase of the
number of clusters, and a decrease of the number of voids with
increasing $D_t$ (for smoothing length 1).

The presence of weak filaments and sheets in the full DM model L512.00
is expressed also in the filling factor and length functions.  At
small threshold densities the single high-density region (cluster)
fills almost the whole sample volume, and the volume of the largest
void is very small. Largest clusters percolate at threshold density
$D_t = 2 - 5$,  largest voids percolate at $D_t = 0.1 - 0.4$,
depending on the smoothing scale.

\begin{figure*}  
\centering 
\hspace{2mm}  
\resizebox{0.45\textwidth}{!}{\includegraphics*{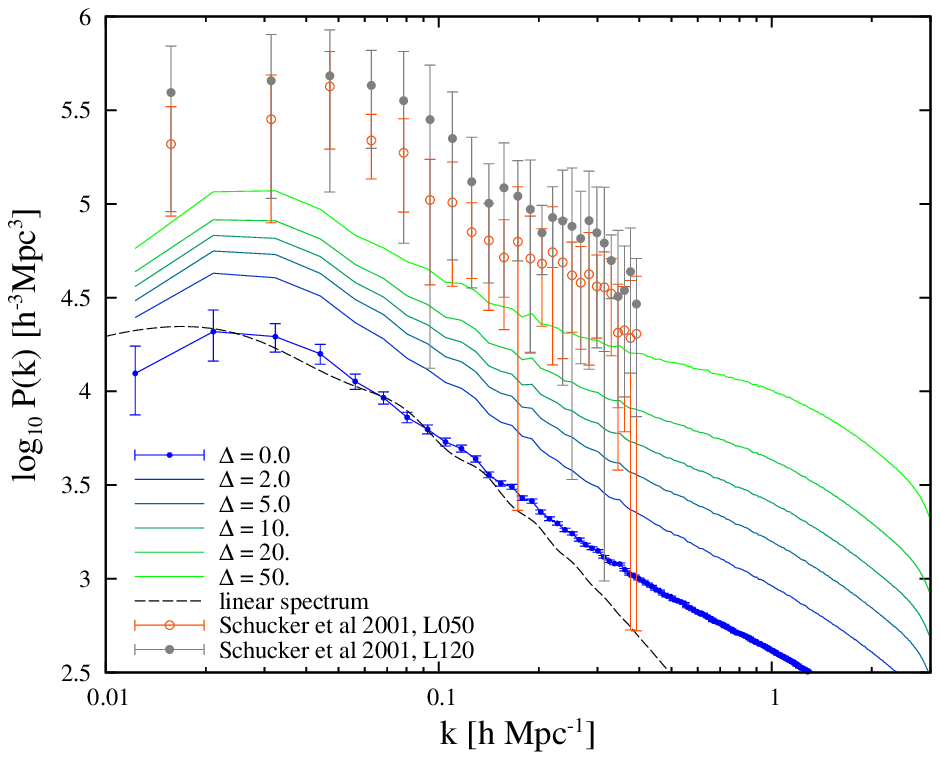}}
\hspace{2mm}  
\resizebox{0.45\textwidth}{!}{\includegraphics*{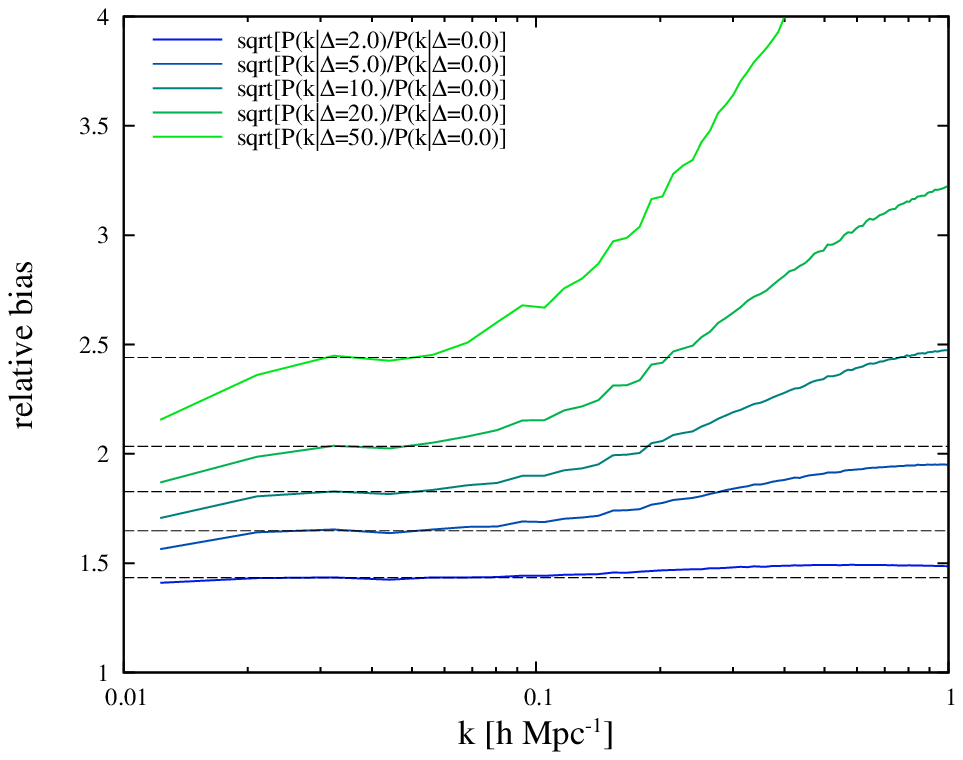}}\\
\hspace{2mm}  
\caption{{\em Left:} Power spectra of particle density limited L512
  model spectra.  With open and filled symbols we show power spectra
  of  X-ray  flux-limited samples  L050 and L120  of the REFLEX
  survey by \citet{Schuecker:2001ly}.  For comparison
  the linear power spectrum is shown. {\em Right}: relative bias
  functions for power spectra of density limited L512 models.  }
\label{fig:bias} 
\end{figure*}

In simulated and real galaxy samples  low-density DM filaments and sheets 
are not present.  This leads to important differences in percolation
functions. First of all, voids are connected (percolated) at all
threshold densities, thus the length of the largest void is equal to
the size of the sample, and the number of voids is equal to 1.  Only
at very large smoothing kernels some additional voids appear at low
$D_t$, created by excessively  smoothed clusters.  Filling factors of
largest clusters are much lower than for full sample L512.00, and
filling factors of largest voids are higher.  The absence of
connecting filaments between high-density knots of the web makes these
knots isolated systems.  For this reason the number of clusters at low
threshold densities is rather large in all galaxy samples, in
contrast to the sample L512.00.  These small isolated clusters are seen
in Fig.~\ref{fig:Fig11} in the model sample L512.10 and in the observed
sample SDSS.21.  The number of clusters,  $\mathcal{N}(D_t)$, is
almost constant at $D_t \le 2$ for all galaxy  samples L512.10,
SDSS.21 and HR4.12 for high-resolution density fields.  Larger
smoothing restores some filamentary connections between knots, thus
the number of clusters decreases with decreasing $D_t$.   

Notice that all percolation functions of simulated and real galaxy
samples L512.10, SDSS.21 and HR4.12 are qualitatively very
similar. Some minor quantitative differences are mainly due to the
fact that the biased model selection parameter (particle density limit
of the sample L512.10) is not exactly tuned for the best mutual
agreement of percolation functions.  This similarity is remarkable,
since volumes and shapes of SDSS and L512 samples are very different,
as mentioned above.  The insensitivity of percolation functions to
sample volumes, shapes and selection methods is an important property
of the percolation method.  This robustness of percolation functions
has a simple explanation: percolation functions measure the growth of
clusters (and voids) with decreasing threshold limit of the density
field, $D_t$. The size, shape and volume of the largest over-density
region (cluster) and largest under-density region (void) does not
depend on the overall size (volume) and shape of the whole sample.

We conclude that the percolation analysis allows a reliable comparison
on model and observed samples. The percolation analysis shows
quantitatively the presence of large differences between the full DM
dominated model L512.00 on the one side, and three versions of galaxy
samples: L512.10, SDSS.21 and HR4.1 on the other side.

\begin{figure*}  
\centering 
\hspace{2mm}  
\resizebox{0.45\textwidth}{!}{\includegraphics*{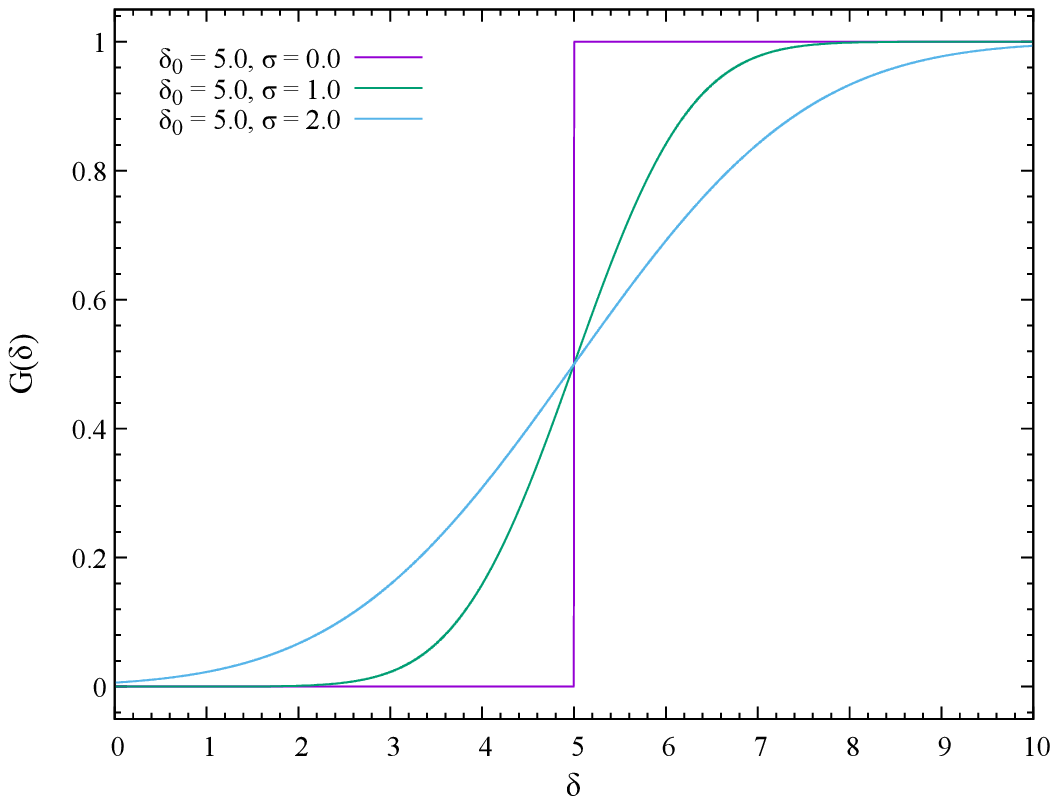}}
\hspace{2mm}  
\resizebox{0.45\textwidth}{!}{\includegraphics*{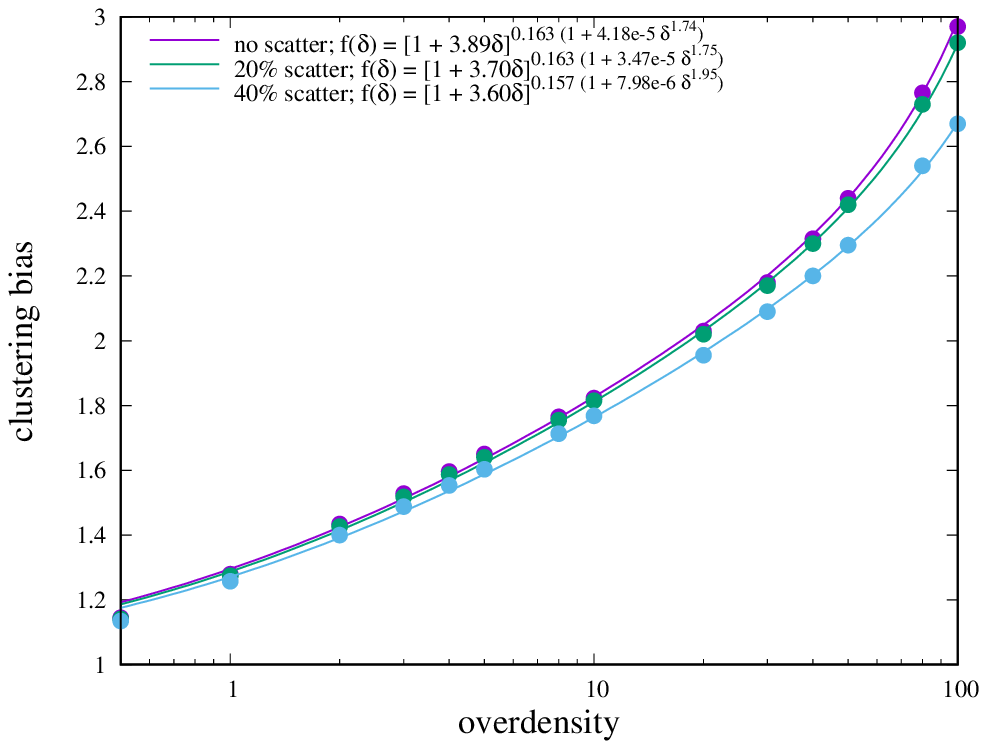}}\\
\caption{{\em Left:} The probability function $G(\delta)$ for 
   L512.5 model with particle density limit
  $\delta_0=5$ and scatter of the particle density limit $\sigma=0,~1,~2$. 
  {\em Right:} bias parameters $b$ as functions of the particle density limit,
  $\delta_0$, found from power spectra of biased models for particle
  density limits, $\delta_0$. Coloured filled circles show actually
  calculated bias parameters from spectra for three values of the
  scatter of the particle density limit, coloured curves show bias
  functions according to Eq.~(\ref{gert}). }
\label{fig:Fig9} 
\end{figure*}

\subsection{Recipes to select galaxy samples}

Galaxy samples were selected for the percolation and power spectrum
analyses using three recipes: halo mass limits, galaxy luminosity
limits and particle local density limits.  In principle we could use a
simulated galaxy model, where halos are filled with galaxies using the
halo occupation distribution (HOD), as done by \citet{Tinker:2009bh}
in the study of the void phenomenon.  However, such synthetic galaxies
are generated only inside DM halos, thus this model is actually a
variant of the HR4 model.  Comparison of galaxy and matter samples was
made using respective density fields.  The difference in the
calculation of respective density fields lies in the use of spatial
coordinates.  In halo mass limited samples only positions of halo
centres were used. Large halos host many galaxies and have dimensions
exceeding the resolution scale of density fields, 1~\Mpc.  Thus
density fields, calculated from halo data, are more rough, what is
seen also in high-resolution density field maps. If data on locations
of individual galaxies are used, as in the case of SDSS samples, we
get a more smooth density field.  When positions of all individual
particles are used, as in the case of L512 model based density fields,
we get an even more smooth density field. However, differences in
density fields are small (see Fig.~\ref{fig:Fig11}), and have little
influence to percolation functions and the distribution of
densities. For this reason percolation functions and density
distributions of real and simulated galaxies are qualitatively very
similar, as seen in Figs.~\ref{fig:Fig14}, \ref{fig:Fig1},  \ref{fig:Fig2},  and \ref{fig:FigNum}.

\section{Power spectra and bias parameters of models}

In this Section we calculate power spectra of biased L512 model
samples.  We investigate the influence of fuzzy particle density
limit.  Our goal is to find the bias as a function of the particle
density limit $\delta_0$, used in the selection of biased model
samples.

\subsection{Power spectra of biased  L512 models}

We calculated power spectra of full L512.00 model with all DM
particles included, $P_m = \Delta_m^2(k)$, and for biased model L512.i
samples (clustered particle samples),
$P_C(k,\delta_0) = \Delta_C^2(k,\delta_0)$, using particle density
limits, $\delta_0$, according to Table~\ref{Tab2}. We applied standard
procedures, used to calculate power spectra of numerical simulations
\citep{Eisenstein:1999fk}.  Power spectra for the full model L512.00
and for biased models L512.i, are shown in left panels of
Fig.~\ref{fig:bias}.  The blue solid line with error ticks is for the
spectrum of the model with all matter, L512.00 (particle density limit
$\delta_0=0$).  Black dashed line shows the linear power spectrum for
this model. Thin coloured lines are for power spectra, calculated for
biased L512.i models according to Table~\ref{Tab2}.  Symbols with
error bars are power spectra of flux-limited X-ray selected clusters
of galaxies, samples L050 and L120, according to
\citet{Schuecker:2001ly}.

Right panel of Fig.~\ref{fig:bias} shows relative bias functions
$b_C(k,\delta_0)=\sqrt{P_C(k,\delta_0)/P_m(k)}$.  Bias functions
depends on the wavenumber $k$.  They have a plateau around
$k \approx 0.03$~$h$~Mpc$^{-1}$, similar to the plateau found by
\citet{Percival:2001aa}.  We used the plateau at
$k \approx 0.03$~$h$~Mpc$^{-1}$ to find bias parameters for model
samples as a function of particle density limits.  Results are seen as
dark circles in the upper curve on the right panel of
Fig.~\ref{fig:Fig9}.

\begin{figure*}
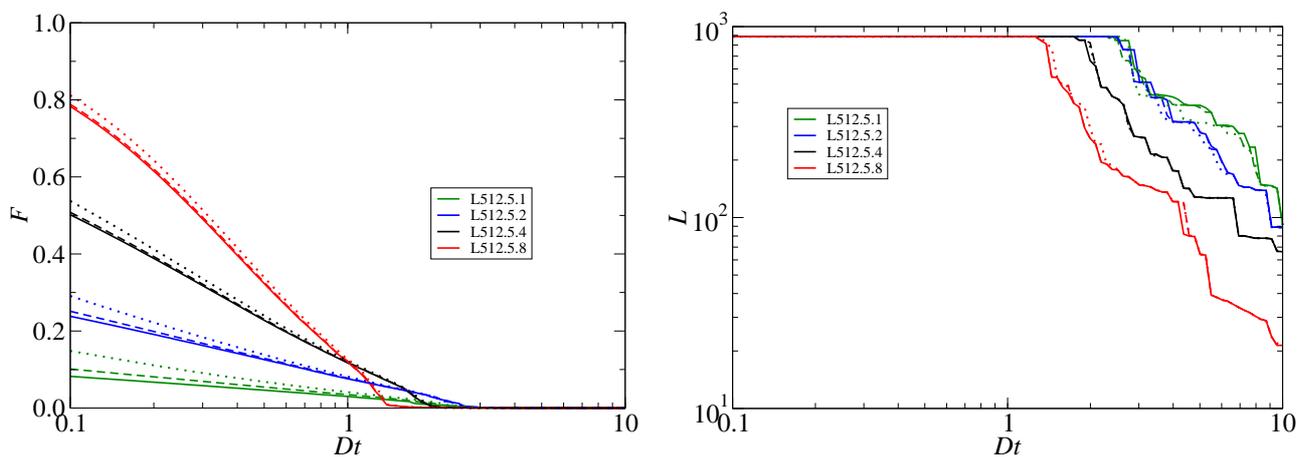
  
\centering 
\hspace{2mm}  
\resizebox{0.45\textwidth}{!}{\includegraphics*{L512Bias5_Vmax-D0_new2.eps}}
\hspace{2mm}  
\resizebox{0.45\textwidth}{!}{\includegraphics*{L512Bias5_Lmax-D0_new2.eps}}\\
\hspace{2mm}  
\caption{
  Percolation functions using fuzzy particle density limits of
  the model L512.5 with $\delta_0=5$.  {\em Left:} filling factor
  functions, {\em right}: length functions of largest superclusters.
  Solid lines are for functions with fuzziness dispersion $\sigma=0$,
  i.e. sharp particle density limit, dashed lines are for dispersion
  $\sigma=1$ and dotted lines for $\sigma=2$.}
\label{fig:fuzzy} 
\end{figure*}

\subsection{Influence of a fuzzy particle density limit}

The use of the local density $\delta_0$ as the only parameter to determine
the fate of particles in the web is a simplification.  Actually
particles of slightly variable densities can end up in  halo-type  density
enhancements, which can be considered as  model equivalents to  real
galaxies. For this reason the density limit to divide particles into
high- and low-density populations is fuzzy. 

To find the influence of a fuzzy particle density limit we made the
power spectrum and percolation function analyses for the L512 model,
using a series of particle local density limits.  We selected particles for
this model as follows: particle with local densities $x \le x_{min}$
are excluded, particles with densities $x$ within limits
$x_{min} \dots x_{max}$ are included with a probability $G(x)$, which
depends on the location within the window, and particles with
densities $x > x_{max}$ are all included.  Here we use the designation
$x=D(\mathbf{x})/D_m$, where $D(\mathbf{x})$ is the
density at location $\mathbf{x}$, and $D_m$ is the mean density. 
Limits $x_{min}$ and
$x_{max}$ are determined by the choice  of the fuzziness distribution
function $p(x)$.  The probability function $G(x)$ to include particles
to the clustered population was
calculated applying the standard procedure:
\begin{equation}
  G(x) = \int_0^{x}  p(x)dx.
\end{equation}
For the fuzziness distribution function  we used the normal distribution
\begin{equation}
  p(x)dx = 
  \frac{1}{\sigma\sqrt{2\pi}} \exp[-\frac{(x - x_0)^2}{2\sigma^2}]dx,
\end{equation}
where $x_0 = \delta_0$ is the location of the sharp limit, and the
dispersion of the particle density limit $\sigma$ depends on the location of the limit, 
$\sigma= f\times \delta_{0}$;   $f$ is the fuzziness parameter.

Power spectra were calculated for a series of particle density limits,
$\delta_{0}$, using fuzziness parameter values $f=0.1,~0.2,~04$. The
limit $f=0$ corresponds to a sharp particle density limit at
$\delta_0$.  Probability distributions $G(\delta)$ are shown in the
left panel of Fig.~\ref{fig:Fig9} for the particle density limit
$\delta_0 = 5$.  It should be noticed that the form of the function
$G(\delta)$ is the same for all particle density limit values, since
the dispersion $\sigma$ is proportional to the density limit,
$\delta_{0}$.  Using power spectra we calculated the clustering bias
parameter $b$, shown by coloured filled symbols in the right panel of
Fig.~\ref{fig:Fig9}.

The bias parameter as function of the particle density limit can be
approximated as follows:
\begin{equation}
b(\delta_0) = (1 + a~\delta_0)^{c(1+d~\delta_0^n)}.
\label{gert}
\end{equation}
This approximation gives a good fit of biased L512 models within particle density limits
$0.5 \le \delta_0 \le 100$, see coloured lines in Fig.~\ref{fig:Fig9}.  
Parameters [a,~c,~d,~n] depend on the fuzziness parameter $f$, used in
calculations of power spectra.  For $f=0$ and $f=0.1$ parameters have
values [a,~c,~d,~n]  = [3.89,~0.163,~4.18E-5,~1.74], for $f=0.2$ we get
[a,~c,~d,~n]  = [3.70,~0.163,~3.47E-5,~1.75], and for $f=0.4$ we have
[a,~c,~d,~n]  = [3.60,~0.157,~7.98E-6,~1.95]. 

We calculated density fields and made the percolation analysis for the
model L512.5, using the particle density limit $\delta_0=5$, the
scatter of particle density limit $\sigma = 1,~2$, and applying
smoothing kernels $R_B= 1,~2,~4,~8$~\Mpc.  Percolation functions
$F(D_t)$ and $L(D_t)$ are shown in Fig.~\ref{fig:fuzzy}.  Solid lines
are for functions using sharp particle density limits with
$\sigma = 0$, dashed lines using particle density limit scatter
$\sigma = 1$, and dotted lines with scatter $\sigma = 2$, which
correspond to fuziness parameter values $f=0.0,~0.2,~0.4$,
respectively.  Our results indicate that the scatter $\sigma = 1$
yields percolation functions, almost identical to percolation
functions with a sharp particle density limit, $\sigma = 0$.  Even the
very large scatter $\sigma = 2$ causes very small changes to
percolation functions.  These calculations showed that power spectra
and percolation functions for fuzzy particle density limit are close
to results obtained with a sharp limit.  Near the low particle density
limit $\delta_0$ the fraction of matter in high-density regions,
$F_C$, determines the effective biasing factor, as shown already by
\citet{Einasto:1999ku}.  Thus we consider the particle density limit
$\delta_0$ as an effective limit.

The bias function,
\begin{equation}
b_C(\delta_0)=\sqrt{P_C(0.03,\delta_0)/P_m(0.03)},
\label{bias}
\end{equation}
is given in Table~\ref{Tab2} and in Fig.~\ref{fig:Fig9}.  It was calculated using the plateau at
$k=0.03$~Mpc$^{-1}$ and applying a smoothed value according to
Eq.~(\ref{gert}).  The bias function yields an one-to-one relationship
between the bias parameter $b$ and the particle selection parameter
$\delta_0$ of biased models.  The bias function depends on the
cosmological parameters of the model, but only weakly, since we use
ratios of power spectra of the same model.  It depends also on the
particle selection method. The fuzziness
analysis shows that the selection method is rather robust.

\section{Percolation analysis of simulated and real galaxy fields}
 
In this Section we continue the extended percolation analysis of
observed and simulated clusters.  We compare percolation functions for
different limiting parameters (luminosities of galaxies, particle
density limits, masses of halos), and for different samples ---
observed vs. simulated samples.  Our goal is to find density limits
$\delta_0$ of biased models, which correspond to luminosity limited
SDSS samples, and mass limited HR4 samples.

\begin{figure*}
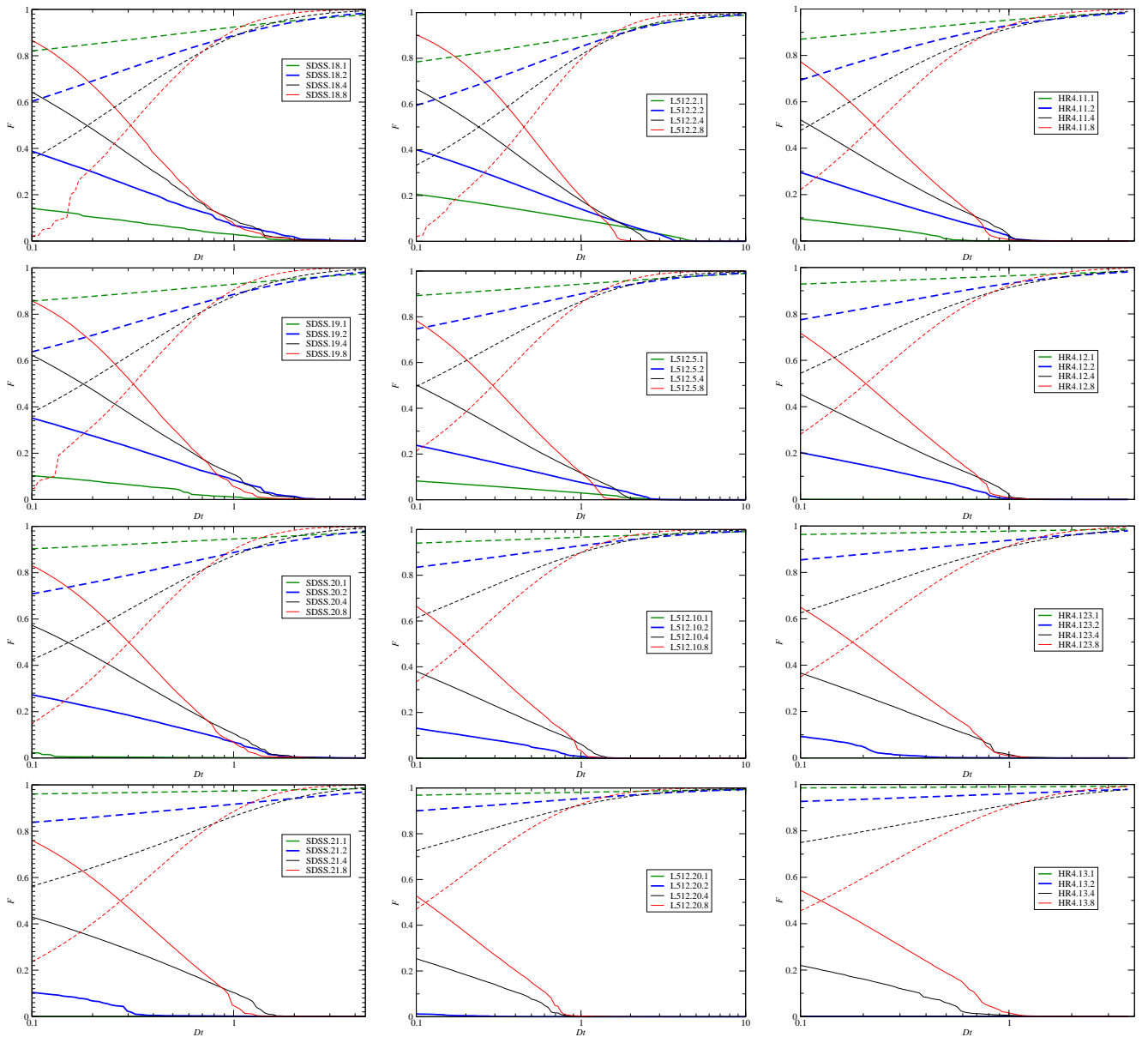
  
\centering 
\hspace{2mm}  
\resizebox{0.30\textwidth}{!}{\includegraphics*{dr10_m18_Vmax-D0_bias45.eps}}
\hspace{2mm} 
\resizebox{0.30\textwidth}{!}{\includegraphics*{L512Bias2_Vmax-D0_4.eps}}
\hspace{2mm} 
\resizebox{0.30\textwidth}{!}{\includegraphics*{HR4.11_Vmax-D0_new.eps}}\\
\hspace{2mm}  
\resizebox{0.30\textwidth}{!}{\includegraphics*{dr10_m19_Vmax-D0_bias45.eps}}
\hspace{2mm} 
 \resizebox{0.30\textwidth}{!}{\includegraphics*{L512Bias5_Vmax-D0_4.eps}}
  \hspace{2mm}
\resizebox{0.30\textwidth}{!}{\includegraphics*{HR4.12_Vmax-D0_new.eps}}\\
\hspace{2mm}  
\resizebox{0.30\textwidth}{!}{\includegraphics*{dr10_m20_Vmax-D0_bias45.eps}}
\hspace{2mm} 
 \resizebox{0.30\textwidth}{!}{\includegraphics*{L512Bias10_Vmax-D0_4.eps}}
\hspace{2mm}  
\resizebox{0.30\textwidth}{!}{\includegraphics*{HR4.123_Vmax-D0_new.eps}}\\
\hspace{2mm}  
\resizebox{0.30\textwidth}{!}{\includegraphics*{dr10_m21_Vmax-D0_bias45.eps}}
\hspace{2mm} 
  \resizebox{0.30\textwidth}{!}{\includegraphics*{L512Bias20_Vmax-D0_4.eps}}
\hspace{2mm}  
\resizebox{0.30\textwidth}{!}{\includegraphics*{HR4.13_Vmax-D0_new.eps}}\\
\hspace{2mm}  
\caption{Filling factor functions of largest clusters and voids,
  $\mathcal{F}(D_t) = V_{\mathrm{max}}/V_0$, as functions of the
  threshold density, $D_t$.  {\em Left panels:} luminosity limited
  SDSS samples, {\em middle panels} particle density limited (biased) L512
  model samples, {\em right panels:} HR4 model samples.  Left panels
  from {\em top} to {\em bottom} are for luminosity limits
  $M_r - 5\log h = -18.0,~-19.0,~-20,0,~-21.0$; middle panels from {\em top} to
  {\em bottom} are for samples with particle density limit
  $\delta_0=2,~5,~10,~20$; right panels from {\em top} to {\em bottom}
  are for halo mass limited samples HR4.11, HR4.12, HR4.123, HR4.13.
  Functions for clusters are plotted with solid lines, and for voids with
  dashed lines. }
\label{fig:Fig1} 
\end{figure*}

\begin{figure*}  
\centering 
\hspace{2mm}  
 \resizebox{0.30\textwidth}{!}{\includegraphics*{dr10_m18_Lmax-D0_bias45.eps}}
\hspace{2mm} 
 \resizebox{0.30\textwidth}{!}{\includegraphics*{L512Bias2_Lmax-D0_new.eps}}
\hspace{2mm} 
\resizebox{0.30\textwidth}{!}{\includegraphics*{HR4.11_Lmax-D0_new.eps}}\\
\hspace{2mm}  
  \resizebox{0.30\textwidth}{!}{\includegraphics*{dr10_m19_Lmax-D0_bias45.eps}}
\hspace{2mm} 
\resizebox{0.30\textwidth}{!}{\includegraphics*{L512Bias5_Lmax-D0_new.eps}}
  \hspace{2mm}
\resizebox{0.30\textwidth}{!}{\includegraphics*{HR4.12_Lmax-D0_new.eps}}\\
  \hspace{2mm}
\resizebox{0.30\textwidth}{!}{\includegraphics*{dr10_m20_Lmax-D0_bias45.eps}}
\hspace{2mm} 
 \resizebox{0.30\textwidth}{!}{\includegraphics*{L512Bias10_Lmax-D0_new.eps}}
  \hspace{2mm}
\resizebox{0.30\textwidth}{!}{\includegraphics*{HR4.123_Lmax-D0_new.eps}}\\
  \hspace{2mm}
\resizebox{0.30\textwidth}{!}{\includegraphics*{dr10_m21_Lmax-D0_bias45.eps}}
\hspace{2mm} 
 \resizebox{0.30\textwidth}{!}{\includegraphics*{L512Bias20_Lmax-D0_new.eps}}
  \hspace{2mm}
\resizebox{0.30\textwidth}{!}{\includegraphics*{HR4.13_Lmax-D0_new.eps}}\\
  \hspace{2mm}
  \caption{Length functions of largest clusters,
    ${L}(D_t) = L_{\mathrm{max}}$ in \Mpc. Panels and lines as in
    Fig.~\ref{fig:Fig1}.}
\label{fig:Fig2} 
\end{figure*}

\begin{figure*}
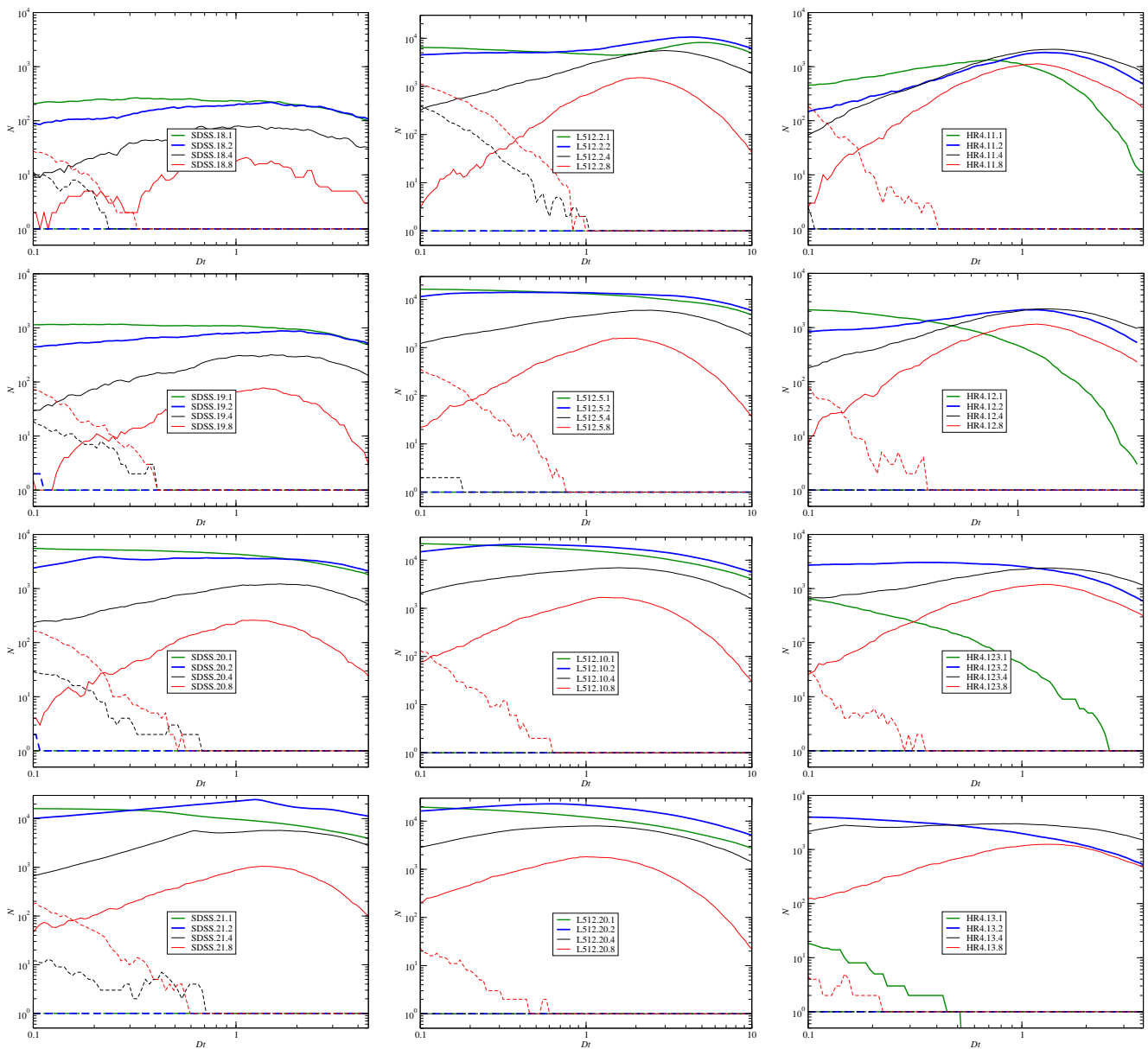
  
\centering 
\hspace{2mm}  
 \resizebox{0.30\textwidth}{!}{\includegraphics*{dr10_m18_N-D0_bias45.eps}}
\hspace{2mm} 
 \resizebox{0.30\textwidth}{!}{\includegraphics*{L512Bias2_N-D0_4.eps}}
\hspace{2mm} 
\resizebox{0.30\textwidth}{!}{\includegraphics*{HR4.11_N-D0_new.eps}}\\
\hspace{2mm}  
  \resizebox{0.30\textwidth}{!}{\includegraphics*{dr10_m19_N-D0_bias45.eps}}
\hspace{2mm} 
\resizebox{0.30\textwidth}{!}{\includegraphics*{L512Bias5_N-D0_new.eps}}
  \hspace{2mm}
\resizebox{0.30\textwidth}{!}{\includegraphics*{HR4.12_N-D0_new.eps}}\\
  \hspace{2mm}
\resizebox{0.30\textwidth}{!}{\includegraphics*{dr10_m20_N-D0_bias45.eps}}
\hspace{2mm} 
 \resizebox{0.30\textwidth}{!}{\includegraphics*{L512Bias10_N-D0_new.eps}}
  \hspace{2mm}
\resizebox{0.30\textwidth}{!}{\includegraphics*{HR4.123_N-D0_new.eps}}\\
  \hspace{2mm}
\resizebox{0.30\textwidth}{!}{\includegraphics*{dr10_m21_N-D0_bias45.eps}}
\hspace{2mm} 
 \resizebox{0.30\textwidth}{!}{\includegraphics*{L512Bias20_N-D0_4.eps}}
  \hspace{2mm}
\resizebox{0.30\textwidth}{!}{\includegraphics*{HR4.13_N-D0_new.eps}}\\
  \hspace{2mm}
  \caption{Number of clusters and voids, $\mathcal{N}(D_t)$.  Number
    functions are calculated with small system exclusion limit
    $N_{\mathrm{lim}}=50$ (L512 and SDSS samples); and with
    $N_{\mathrm{lim}}=500$ (HR4 samples).  Lines as in
    Fig.~\ref{fig:Fig1}.}
\label{fig:FigNum} 
\end{figure*}

\begin{figure*}
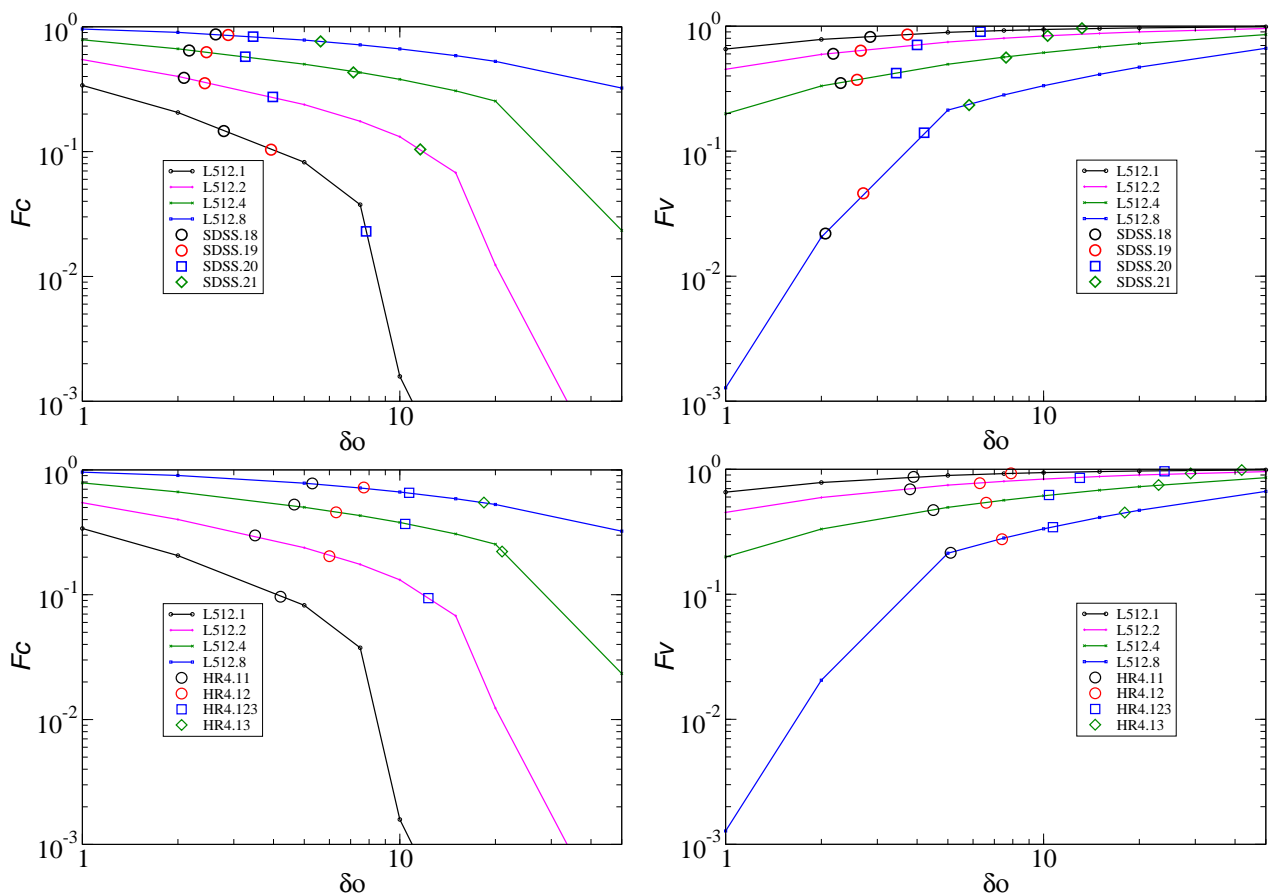
  
\centering 
\hspace{2mm}  
\resizebox{0.44\textwidth}{!}{\includegraphics*{SDSS_log_m18-21_F_Clus-D0_45.eps}}
\hspace{2mm}  
\resizebox{0.44\textwidth}{!}{\includegraphics*{SDSS_log_m18-21_F_Void-D0_45.eps}}\\
\hspace{2mm}  
\resizebox{0.44\textwidth}{!}{\includegraphics*{HR4_11-13_F_C-D0_37.eps}}
\hspace{2mm}  
\resizebox{0.44\textwidth}{!}{\includegraphics*{HR4_11-13_F_V-D0_37.eps}}\\
\hspace{1mm}  
\caption{Filling factor test function of clusters $F_C(\delta_0)$ and voids
  $F_V(\delta_0)$.  Solid lines show test functions of L512 models for
  four smoothing scales, $R_B=1,~2,~4,~8$~\Mpc.  {\em Left panels} are for
  cluster filling factors, {\em right panels} for void filling factors. 
 {\em Upper panels}  show by symbols filling factors of SDSS samples, {\em lower
  panels} filling factors of HR4 samples.  
}
\label{fig:Fig3} 
\end{figure*}

\subsection{Percolation properties of observed and model
  samples}

Percolation properties depend on the smoothing kernel size, since
smoothing makes clusters and connecting filaments larger and helps
clusters to percolate.  We calculated density fields of observed and
model samples, expressing densities in units of the mean density of
particular samples.  The L512 model sample includes all particles of
the model, the HR4 model sample includes all halos, containing at
least 30 particles, the SDSS samples includes all galaxies within the
apparent {\em r} magnitude interval $12.5 \le m_r \le 17.77$.  Using
density fields with these normalisations we calculated all percolation
functions.

The comparison of percolation functions of observed SDSS samples and
HR4 model samples with percolation functions of L512 model samples
showed that percolation functions of SDSS and HR4 samples are shifted
relative to L512 model samples towards higher threshold
densities. This is a well-known effect.  All densities are expressed
in mean density units.  In model samples the mean density includes, in
addition to clustered matter, also DM in low-density regions, where
there are no galaxies, or galaxies are fainter than the magnitude
limit of the observational SDSS survey.  In calculations of the mean
density of observed SDSS samples and HR4 model samples unclustered and
low-density DM is not included.  This means that in the calculation of
densities in mean density units densities are divided to smaller
numbers, which increases density values of SDSS and HR4 samples.

The total number and mass of particles of HR4 model samples is known,
thus it is easy to calculate the density normalisation factor, which
brings density fields to the level, corresponding to all particles.
We found that the fraction of particles in our HR4 samples is
$f=0.37156$.  To bring HR4 density fields to the same normalisation as
L512 density fields, we multiplied all threshold densities by the
factor $f$.  Corrected density thresholds can be used as arguments in
percolation functions.  Since we plot percolation functions using as
argument $\log(D_t)$, the shape of functions does not depend on the
normalisation factor, and we can use the same functions as in our
preliminary analysis, only the argument is shifted.

The comparison of percolation functions of SDSS and HR4 samples shows
that the SDSS sample contains galaxies which correspond to less
massive halos than the HR4 sample. For this reason the correction
factor $f$ for SDSS samples must be closer to unity.  The exact value
of this factor is difficult to calculate. Our analysis of percolation
functions and respective parameters using various density
normalisation factors shows that final results depend on the exact
value of the factor $f$ only rather modestly.  We accepted for SDSS
samples the factor $f=0.45$.

Using corrected density normalisations we show in Fig.~\ref{fig:Fig1}
filling factor functions of largest clusters and voids,
$\mathcal{F}(D_t)=V_{\mathrm{max}}/V_0$,  in Fig.~\ref{fig:Fig2}
length functions ---  lengths of largest clusters,
${L}(D_t) = L_{\mathrm{max}}$ in \Mpc, and in Fig.~\ref{fig:FigNum}
numbers functions $\mathcal{F}(D_t)$.  Length functions of voids are
not defined since voids percolate at all threshold densities.  Left
panels in  Figures are for luminosity limited SDSS samples, middle
panels for particle density limited L512 samples, using particles
density limits $\delta_0=2,~5,~10,~20$, and right panels for halo mass
limited HR4 model samples.  In functions as argument we use the
threshold density, $D_t$, corrected for sample mean density normalisation.
Solid lines in Figures are for clusters, dashed lines for voids;
colours mark different smoothing lengths.

Figs.~\ref{fig:Fig1},  \ref{fig:Fig2} and \ref{fig:FigNum} show that percolation
functions of observed and model samples have surprisingly similar
shapes, and depend in a simple and regular way on sample parameters.
The difference between filling factor functions of observed and model
samples is the largest at lowest threshold density, $D_t=0.1$.  Thus
we shall use  filling factor function values at threshold
density, $D_t=0.1$ as  test parameters to compare samples in a
quantitative manner.  The behaviour of length functions is more
complicated and is discussed later.

\begin{figure*}
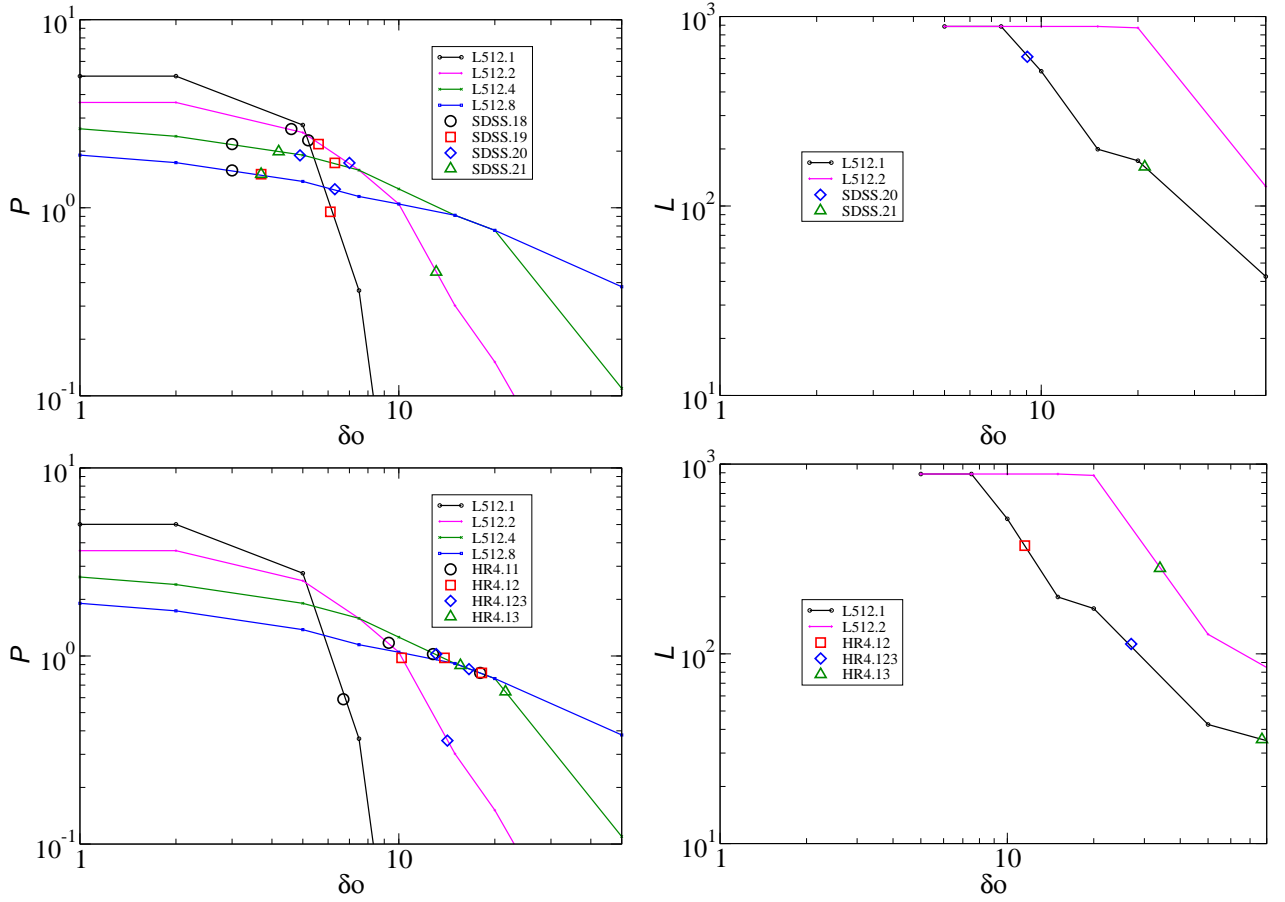
  
\centering 
\hspace{2mm}  
\resizebox{0.44\textwidth}{!}{\includegraphics*{L512_SDSS_Dt_Length-D0.eps}}
\hspace{2mm}  
\resizebox{0.44\textwidth}{!}{\includegraphics*{L512_SDSS_m20-m21_L-D0.eps}}\\
\hspace{2mm}  
\resizebox{0.44\textwidth}{!}{\includegraphics*{L512_HR4_Dt_Length-D0.eps}}
\hspace{2mm}  
\resizebox{0.44\textwidth}{!}{\includegraphics*{L512_HR4_12-13_L-D0.eps}}\\
\caption{ {\em Left} panels shows the cluster percolation threshold
  test, {\em righ} panels show the cluster length test for
  high-luminosity (mass) samples.  {\em Upper} panels are for SDSS
  samples, {\em lower} panels for HR4 samples;.  Solid lines show
  cluster length test functions of L512 models for various smoothing
  scales.  Coloured symbols show SDSS samples of various luminosity
  and HR4 samples of various halo mass limits.  }
\label{fig:Fig4} 
\end{figure*}

\subsection{Filling factor test  of clusters and voids}

The cluster filling factor test parameter is defined as
$F_C = \mathcal{F}_C(0.1)$, and the void filling factor test parameter as
$F_V = \mathcal{F}_V(0.1)$.  Each sample with a different particle
density limit $\delta_0$ and smoothing kernel size $R_B$ yields a
value for $F_C$ and $F_V$.   We calculated these parameters of
the L512 model for a range of particle density limits $\delta_0$, and
define filling factor test functions as follows:
$F_C(\delta_0)$ and $F_V(\delta_0)$. 
Our goal is to find particle density limits $\delta_0$ of L512 models
which correspond to galaxy luminosity limits of SDSS samples and halo mass
limits of HR4 samples.  We shall use the following strategy: we plot filling
factor test functions $F_C(\delta_0)$ and $F_V(\delta_0)$ of biased models L512,
separately for smoothing kernels $R_B=1,~2,~4,~8$~\Mpc.  Results are
shown in  Fig.~\ref{fig:Fig3}.  Solid lines are for
biased L512 models, different colours mark functions calculated using
smoothing scales $R_B=1,~2,~4,~8$~\Mpc.  Left panels are for cluster
filling factors, and right panels for void filling factors.  Filling
factors of SDSS and HR4 samples of various luminosity limits are known, and it
is easy to interpolate, at which particle density limits $\delta_0$
filling factors of SDSS (HR4) samples are equal to filling factors of biased
L512 samples.  This comparison was made separately for each smoothing
kernel value, $R_B$, and for filling factors of clusters and voids.
We tried several interpolation schemes, and found that good results
are obtained by linear interpolation along the $\log\delta_0$ axis.

{\scriptsize 
\begin{table*}[ht]   
\caption{Particle density limit and bias parameters of SDSS and HR4 samples.} 
\label{Tab4}                         
\centering
\begin{tabular}{lcccccc}
\hline  \hline
  Sample   &  $(\delta_0)_F$  &$(\delta_0)_L$& $(\delta_0)_{Ls}$
  &  $b_F$  & $b_L$ & $b_{Ls}$\\
\hline  
(1)&(2)&(3)&(4)&(5)&(6)&(7)\\ 
\hline  
SDSS.18 & $2.5 \pm 0.4$ &$4.9 \pm 1.0$ &$5 \pm 2$
& $1.47\pm 0.03$ & $1.63 \pm 0.05$ & $1.64 \pm 0.11$ \\
SDSS.19 & $3.2 \pm 0.8$ &  $5.9 \pm 1.5$ &$6 \pm 2$ 
& $1.53 \pm 0.06$ & $1.68 \pm 0.07$ & $1.68 \pm 0.09$\\
SDSS.20 &  $5.5 \pm 1.9$ & $8.0 \pm 2.0$ & $10 \pm 3$
& $1.66 \pm 0.09$ & $1.76 \pm 0.07$ & $1.83 \pm 0.09$\\
SDSS.21 & $13.0 \pm 2.9$ & $17 \pm 4$& $20 \pm 5$  
& $1.91 \pm 0.07$ & $1.99 \pm 0.08$ & $2.05 \pm 0.09$\\
\\
HR4.11 & $3.8 \pm 0.4$& $8.0 \pm 1.5$ &$6 \pm 2$
& $1.57 \pm 0.02$&$1.76 \pm 0.05$& $1.68 \pm 0.09$ \\
HR4.12 & $8.0 \pm 2.6$ & $10.8 \pm 2.0$ &$10 \pm 4$
& $1.76 \pm 0.09$&$1.85 \pm 0.06$ & $1.83 \pm 0.12$ \\
HR4.123 & $21 \pm 10$& $21 \pm 5$&$20 \pm 5$
&$2.06 \pm 0.18$&$2.07 \pm 0.09$ & $2.05 \pm 0.09$\\
HR4.13 & $42 \pm 13$ &  $56 \pm 8$&$40 \pm 10$
&$2.35 \pm 0.16$&$2.51 \pm 0.09$ & $2.33 \pm 0.12$\\
\hline 
\end{tabular} \\
\tablefoot{
The columns are:
(1): sample name; 
(2): particle density limit $(\delta_0)_F$, found from filling factor
test;
(3): particle density limit $(\delta_0)_L$, found from length function
test; 
(4): particle density limit $(\delta_0)_{Ls}$, found from length function
shape test; 
(5): bias parameter $b_F$ found from biased L512 spectra using particle
density limits $(\delta_0)_F$ from filling factor test; 
(6): bias parameter $b_L$ estimated from biased L512 spectra 
using particle density limit $(\delta_0)_L$ from cluster length test;
(7): bias parameter $b_{Ls}$ estimated from cluster length test using
particle density limit $(\delta_0)_L$ from cluster length shape test.
}
\end{table*} 
} 

Using this interpolation scheme we found for each SDSS luminosity
limit (HR4 sample halo mass limit) particle density limits $\delta_0$
of biased L512 samples, which correspond to these SDSS (HR4) samples,
separately for clusters and voids, and for different smoothing kernel
length, $R_B$.  Fig.~\ref{fig:Fig3} show that there is a good
agreement between values obtained using clusters and voids, and using
various smoothing kernels.  We consider density fields smoothed with
$R_B=1$ and $R_B=2$~\Mpc\ kernels as the best ones for comparison, and
accept mean values of particle density limits $\delta_0$, found on the
basis of clusters and voids of L512 models for these smoothing
kernels, as representing observed SDSS (HR4) samples of various
luminosity (mass) limits.  We estimated errors of obtained $\delta_0$
values from deviations of individual $\delta_0$ values for clusters
and voids, using smoothing kernels $R_B=1$ and $R_B=2$~\Mpc.  Results
are given in column (2) of Table~\ref{Tab4}.

\subsection{Length function test}

Now we compare SDSS and HR4 samples with reference samples of biased L512 models
using the length function.  For this purpose only cluster length
functions can be used, since voids percolate at all threshold levels,
and void lengths are equal to sample effective lengths.  Most valuable
information comes from the location of length functions for smoothing
kernels $R_B=1$~\Mpc\ and $R_B=2$~\Mpc, in relation to functions for
larger smoothing scales.  But the problem here is that only at low
limiting luminosities (halo masses) clusters percolate, and thus the
percolating density threshold $D_t$ can be found only for these 
samples.

Positions of SDSS and HR4 samples relative to L512 samples can be
found using percolation threshold densities.  Percolation threshold
densities of biased L512 samples as functions of the particle density
limit $\delta_0$ are shown in left panels of Fig.~\ref{fig:Fig4},
separately for various smoothing kernel lengths. Percolation threshold
densities of SDSS and HR4 samples were found using length functions
for largest and second-largest clusters, following
\citet{Klypin:1993aa}.   By interpolation along the $\delta_0$
coordinate we found particle density limits of L512 samples,
corresponding to SDSS and HR4 samples.

\begin{figure*}
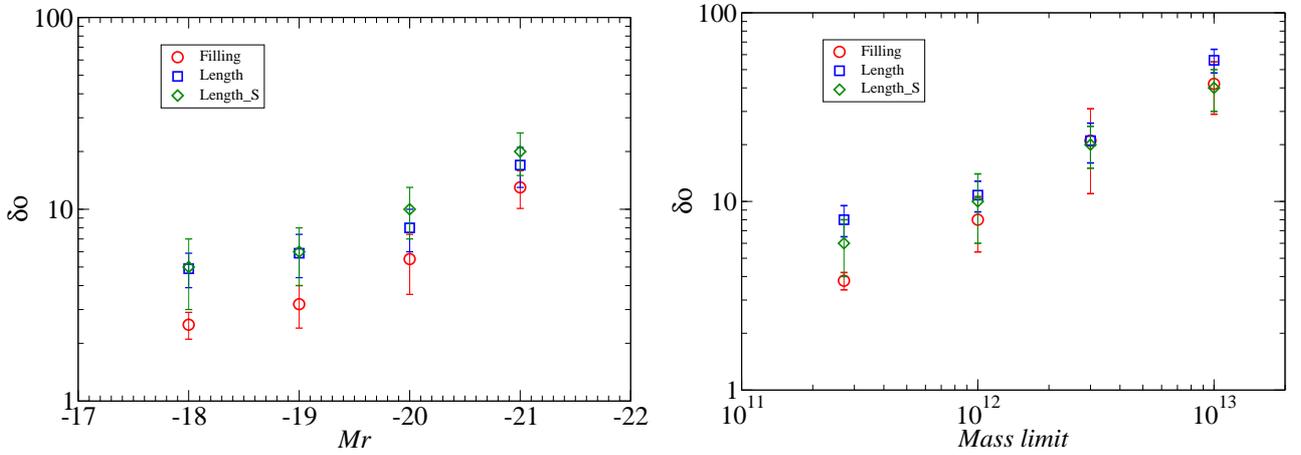
  
\centering 
\hspace{1mm}  
\resizebox{0.46\textwidth}{!}{\includegraphics*{DR10_Do-magnB.eps}}
\hspace{1mm}  
\resizebox{0.44\textwidth}{!}{\includegraphics*{HR4_Do-mass.eps}}\\
\hspace{1mm}  
\caption{Particle density limit parameter $\delta_0$ for SDSS
  luminosity limited samples is in {\em left panel}, and for halo mass
  limited HR4 model samples in {\em right panel}.  Absolute magnitude
  and halo mass limits are used as arguments. Red circles are based on
  filling factor test for clusters and voids, blue boxes from length
  functions test, and green diamonds from length function shape estimates, as
  given in Table~\ref{Tab4}.  }
\label{fig:Fig5} 
\end{figure*}

At high luminosities (halo masses) the largest clusters at small
smoothing lengths are shorter than the total length of the sample, as
seen in plots of density fields of samples L512.5  and SDSS.19 in
Fig.~\ref{fig:Fig11}.  For SDSS and HR4 samples with no percolation we
applied cluster lengths at threshold density $D_t=0.1$, using
interpolation schemes similar to interpolation of filling factor
functions.  This procedure was applied for high luminosity samples
SDSS.20 and SDSS.21, and for high-mass samples HR4.12, HR4.123, and
HR4.13.  Results of this interpolation are shown in right panels of
Fig.~\ref{fig:Fig4}.  We used locations of length functions for
smoothing kernels $R_B=1$ and $R_B=2$~\Mpc.  Errors were calculated
from the scatter of individual values for different smoothing kernels.
Results for particle density limits are given in column (3)
$(\delta_0)_{L}$ of Table~\ref{Tab4} for SDSS and HR4 samples.

\begin{figure*}
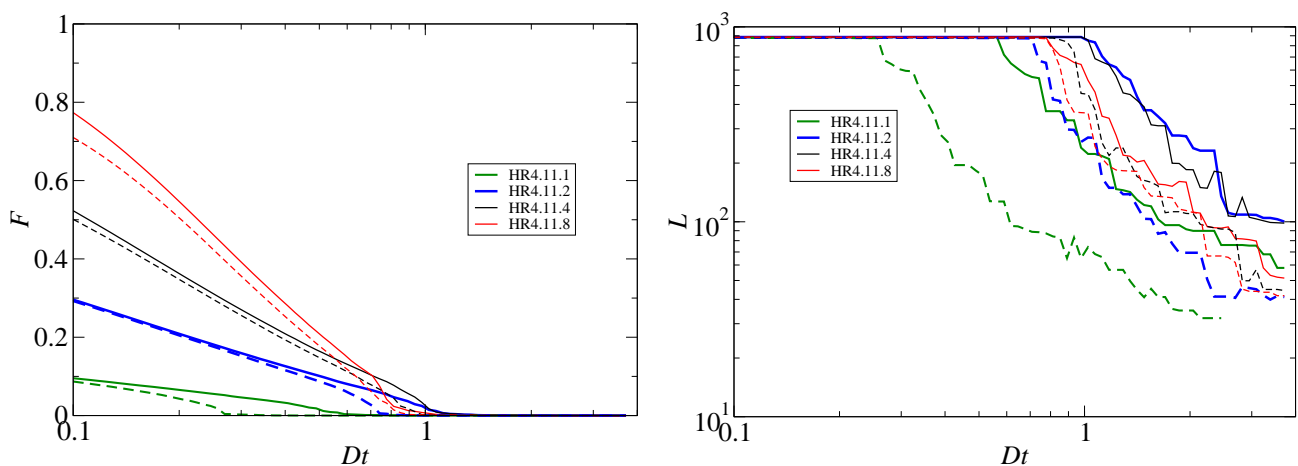
  
\centering 
\hspace{2mm}  
\resizebox{0.45\textwidth}{!}{\includegraphics*{HR4.11_Vmax-D0_fact.eps}}
\hspace{2mm}  
\resizebox{0.45\textwidth}{!}{\includegraphics*{HR4.11_Lmax-D0_fact.eps}}\\
\hspace{2mm}  
\caption{Percolation function tests to check the influence of redshift
  space distortions. {\em 
    Left} panel shows filling factor functions and {\em right}
  panel length functions of HR4.11 models. Solid lines are calculated
  for real space, dashed lines for redshift space. }
\label{fig:Fig10} 
\end{figure*}

One possibility to get simple estimates of positions of SDSS (and HR4)
samples relative to L512 samples is to use the general shape of
cluster length functions.  Here we compare the shape and location of
length functions for smoothing kernels $R_B=1,~2$~\Mpc\ with the shape
and location of length functions for larger smoothing kernels.  For
instance, let us compare the length function of the SDSS.18.1 (here
the second index 1 shows the smoothing scale, $R_B=1$~\Mpc) sample
relative to SDSS.18 samples for larger smoothing kernels, SDSS.18.2,
SDSS.18.4, and SDSS.18.8. Fig.~\ref{fig:Fig2} shows that the length
function of the sample SDSS.18.1 is located between length functions
of samples SDSS.18.4 and SDSS.18.8.  Near the percolation level the
length function of the model sample L512.2.1 lies to the right of
length functions of samples L512.2.2, L512.2.4 and L512.2.8.  But in
the model sample with higher particle density limit L512.5.1 the
length function is located approximately identical to the location of
the length function of the L512.5.2 sample.

Comparing locations of SDSS sample length functions at smoothing
kernels $R_B=1$~\Mpc\ and $R_B=2$~\Mpc\ with locations of respective
L512 samples we find that the shape of the length function of the
SDSS.18 sample corresponds best with the model sample of particle
density limit $\delta_0=5$.  Definitely length functions of the
SDSS.18 samples correspond to model samples L512 within particle
density limits $3 <\delta_0 <7$.  Thus our estimated corresponding
model of the L512 series for the SDSS.18 sample is the model with
$\delta_0 =5 \pm 2$.  This simple length function shape comparison
test uses the information for the whole length function, not only at
percolation or $D_t=0.1$ level.  It was made for all SDSS and HR4
samples, results are given in column (4) $(\delta_0)_{Ls}$ of
Table~\ref{Tab4}.

Length function tests confirm results obtained with the filling factor
test.  As we see from Table~\ref{Tab4}, cluster length tests yield a
bit higher  particle density limits $\delta_0$ 
for SDSS samples than the filling factor test.
Fig.~\ref{fig:Fig5} shows particle density limit parameter
$\delta_0$ and its error for all luminosity limited SDSS samples, and
for halo mass limited HR4 samples, applying filling function and
length function tests.  All tests show that the particle density limit
of biased model samples increases with the increase of the luminosity
(halo mass) of the sample; the increase is regular and has a small
scatter.  Graphs for SDSS and HR4 samples can be brought to a single
diagram if a mass-to-luminosity ratio of SDSS samples,
$M/L_r= 103 \pm 10$, is applied.

\subsection{Redshift space distortions}

Model samples yield true spatial coordinates, observational SDSS
samples are based on angular positions and redshifts.  In using
redshifts to calculate spatial coordinates one must take into account
two types of redshift space distortions: the fingers of God effect,
caused by random motion of galaxies in clusters, and redshift space
distortions (RSD) or the \citet{Kaiser:1987aa} effect, caused by
coherent motions of galaxies towards clusters and superclusters.

The finger of God effect is taken into account in catalogues of groups
and clusters, used in calculations of the SDSS density field
\citep{Tempel:2014uq}.  The Kaiser effect leads to an apparent
flattening of the structure --- pancakes of God. For this reason the
SDSS density field is distorted. Filamentary connections between
over-density regions are disturbed by the Kaiser effect.  Filling
factor test functions are shown in Fig.~\ref{fig:Fig3}, cluster length
test functions in Fig. ~\ref{fig:Fig4}.  In these Figures filling
factors and lengths of L512 model samples represent true model filling
factors and lengths.  To correct for this effect points corresponding
to SDSS samples must be shifted, which leads to changes of particle
density limits $\delta_0$.  The problem is, How much?

In answering this question the comparison of filling factor and length
tests of HR4 model samples is of help.  The analysis described above was based on
true positions of halos in real space.  We made a new analysis, using positions and
velocities of halos of the HR4.11 sample to a calculate a version of
the sample in redshift space.  Positions of all halos were shifted
from real to redshift space for an observer, located outside of the
box by 100~\Mpc, at the centre below the $x=y=0$ plane.  We calculated
percolation functions for both versions of the HR4.11 sample.  Results
are shown in top panels of Fig.~\ref{fig:Fig10}, the left panel is for 
filling factor functions, and the right panel for  length functions.

\begin{figure*}
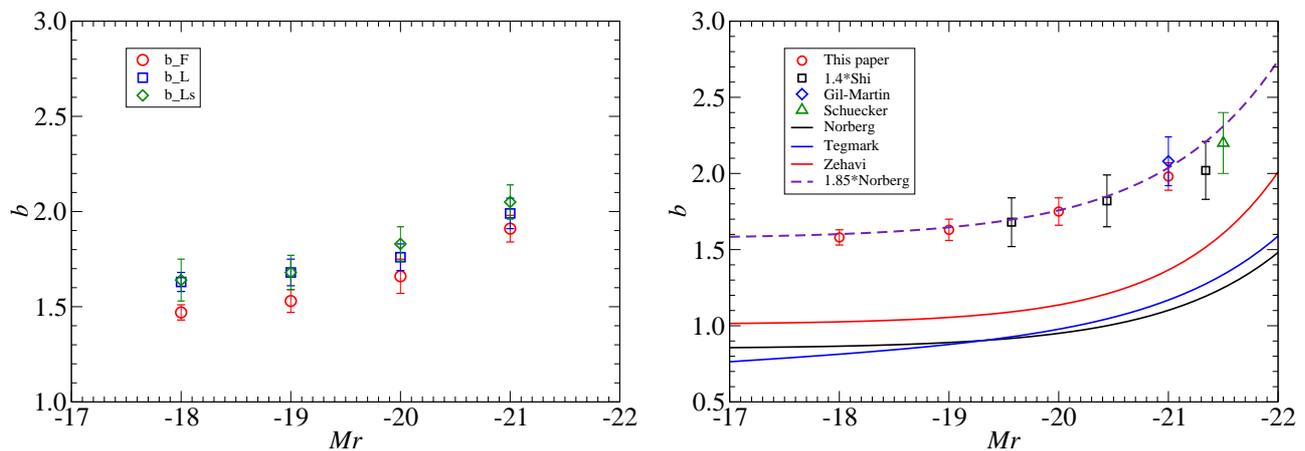
  
\centering 
\hspace{2mm}  
\resizebox{0.45\textwidth}{!}{\includegraphics*{biasmodelN20.eps}}
\hspace{2mm}  
\resizebox{0.45\textwidth}{!}{\includegraphics*{biasmodelN30B.eps}}\\
  \caption{ {\em Left panel:} bias parameter $b$ values as function of the
    absolute magnitude limit $M_r - 5\log h $ of the sample.  Red circles show
    bias parameter values as found from biased L512 spectra using the
    filling factor test, $b_F$, blue boxes indicate bias parameters as
    found from spectra using the length function test, $b_L$, and
    green diamonds show bias parameters using the length function
    general shape test $b_{Ls}$.
    {\em Right panel:} Bias function found in this paper compared with bias
  functions found in other studies. Bias parameter values for SDSS
  galaxies of various absolute magnitude limits found in this paper
  are marked by red circles with error bars. Bias parameter values of
  SDSS galaxies found by \citet{Shi:2018bh} and applying normalising
  factor $b_\circ=1.4$ are marked by black squares with error bars;
  bias parameter for BOSS galaxies according to
  \citet{Gil-Marin:2017ly} by blue diamond, and for X-ray clusters of
  galaxies by \citet{Schuecker:2001ly} by green triangle.  Dashed 
  line shows the fit by \citet{Norberg:2001aa}, applying bias
  normalising factor $b_\circ=~1.85$. Black, blue and red lines show fits by
 \citet{Norberg:2001aa},  \citet{Tegmark:2004aa} and
 \citet{Zehavi:2011aa}. 
    }
\label{fig:bias1} 
\end{figure*} 

We used filling factors at threshold density level $D_t=0.1$ to find
particle density limits of L512 samples, corresponding to HR4 models.
This test shows, that for smoothing lengths 1 and 2~\Mpc, HR4.11
models in real and redshift space yield practically identical
$\delta_0$ limits.  In the cluster percolation threshold (length) test HR4.11
samples in redshift space yield a mean value $\delta_0 = 9.6 \pm 1.9$,
which is a bit larger than for real space, $\delta_0 = 8.0 \pm 1.5$,
see Table~\ref{Tab4}.  If such relative corrections are applied to
observed samples, then particle density limits $(\delta_0)_L$ in real
space would be $4.1,~4.9,~6.7,~14.1$ for samples SDSS.18, SDSS.19,
SDSS.20 and SDSS.21, respectively.  Such corrections bring particle
density limits $(\delta_0)_L$ closer to limits, found from the filling
factor test, $(\delta_0)_F$.

The length function test compares essentially percolation length of
samples, the filling factor test the volumes of clusters and voids at
very low threshold density $D_t=0.1$.  Our check using HR4 samples
shows that differences found from the comparison of results for real
and redshift space are small.  Filling factor and length tests
represent very different properties of the cosmic web; it is
surprising that both tests yield so close results.  We can make a
tentative conclusion that  redshift space distortion effect is
small, and that differences of filling factor and length tests of SDSS
samples are mainly due to random factors.

\subsection{Luminosity dependence}

We applied the function (\ref{gert}) with parameter set for $f=0$ to
find bias parameter values of biased L512 models, $b_F$,~$b_L$ and
$b_{Ls}$, corresponding to SDSS and HR4 samples.  Results are given in
Table~\ref{Tab4}.  Bias parameter values of SDSS samples are shown in
left panel of Fig.~\ref{fig:bias1} as function of the absolute
magnitude limit $M_r$ of samples.  Red circles show $b_F$, calculated
from L512 spectra for biased models using filling factor test; blue
and green symbols show $b_L$and $b_{Ls}$, found from spectra of biased
L512 models using two length function tests.  Errors were found from
the mean scatter of $b_F$, $b_L$ and $b_{Ls}$ values for luminosity
limited SDSS samples, as shown in Table~\ref{Tab4}.  We accept an
arithmetic mean of our three tests, $b_F$, $b_L$, $B_{Ls}$, shown by
red circles in the right panel of Fig. ~\ref{fig:bias1}.  The accepted
error corresponds to a characteristic error from one measurement.  It
is internal error --- possible systematic errors due to the method
must be discussed separately. 

Available data indicate that the relative bias function
$f(L) =b(L)/b_\circ$ is approximately constant at low luminosities
$L < L_\ast$ at level $f(L<L_\ast)) \approx 0.9$, which leads to
$b(L < L_\ast) \approx 0.9 \times b_\circ$.  The flattening of the
$b(L)$ relation at low luminosities $L$ is probably due to the
nature of the distribution of faint galaxies.  As shown by
\citet{Tempel:2009sp}, first ranked galaxies have a tendency of cutoff
at magnitudes $M - 5\log_{10} h \approx -17$ in photometric system of
the 2dF survey $b_J$. Satellite galaxies can have fainter
luminosities, but satellites are located only around main galaxies
\citep{Einasto:1974b}.  Thus power spectra and percolation properties
of very faint galaxies cannot be very different from properties of
galaxies corresponding to the faint end of the luminosity function of
ventral  galaxies.

The luminosity dependence of our data are very well fit by
\citet{Norberg:2001aa} bias function
\begin{equation}
f(>L)=b(>L)/b_\circ = 0.85 + 0.15(L/L_\ast), 
\label{bias0}
\end{equation}
where $L$ is the luminosity limit of galaxies of the sample, $L_\ast$
is the characteristic luminosity of the sample
\citep{Schechter:1976fk}, and $b_\circ$ is the bias normalising
factor.  Fitting our bias parameters for SDSS samples to the 
 \citet{Norberg:2001aa} bias function yields a 
normalising factor $b_\circ=1.85 \pm 0.15$, and we get for the
luminosity dependence of SDSS galaxies:
\begin{equation}
b(>L) = 1.85 \times (0.85 + 0.15(L/L_\ast)). 
\label{bias000}
\end{equation}
The error was found from
the scatter of tests for $b_F$, $b_L$, $B_{Ls}$.
\citet{Norberg:2001aa} defined the relative bias function so
that $f(L_\ast)=1$.  Thus our analysis suggests for the bias parameter
of $L_\ast$ galaxies is $b_\ast = 1.85 \pm 0.15$.

Our bias function is based on the comparison of power spectra of
biased model samples L512.i and full model sample L512.00.  Since the
comparison is differential, possible errors in cosmological parameters
of the model are minimal.  The critical element of the method is the
sharp density limit $\delta_0$ used to select particles for biased
samples. We checked the influence of sharpness of the density limit
using fuzzy limits.  However, there remains the question: How well
such biased model samples represent luminosity limited SDSS galaxy
samples?  It is possible that sharp particle density limit and sharp
SDSS galaxy luminosity limit yield slightly different samples near
lower borders of limits.

We can check possible differences using number functions.
Fig.~\ref{fig:FigNum} shows that at low and medium threshold densities
$0.1 \le D_t \le 1$ and small smoothing lengths number functions of
all biased model samples and luminosity limited SDSS samples are
almost flat.  In this threshold density interval samples are dominated
by small isolated clusters, practically identical for the same sample
at various threshold density levels, thus the number of clusters
remains the same.  In samples with larger smoothing length the number
of clusters at low $D_t$ is smaller, because smoothing joins these
isolated clusters at low threshold limit to filaments, and isolated
clusters almost disappear.  At threshold density $D_t=0.1$ and
smoothing length $R_B=8$~\Mpc\ samples SDSS.18 and L512.2 contain only
a few isolated clusters due to faint galaxy filaments connecting knots
to single objects.  In samples of higher limiting luminosity or
particle density limit the number of clusters at $D_t=0.1$ gets
larger, since filaments connecting knots became invisible, see
Fig.~\ref{fig:FigNum}.  In this respect simulated and real galaxy
samples yield qualitatively very similar results.  We may conclude
that our method to find biased model samples using sharp particle
density limits is a fair representative to SDSS luminosity limited
samples.

As discussed above, redshift space distortions can increase length
function test parameters $(\delta_0)_L$ and corresponding bias
parameters $b_L$.  Thus, if one prefers the filling factor test, one
can use for the bias parameter of $L_\ast$ galaxies a value,
$b_\ast = 1.70 \pm 0.15$.  This does not influence our main conclusion
that the bias parameter of $L_\ast$ galaxies is $b_\ast \gg 1.0$.

\subsection{Comparison with other data}

In the right panel of Fig.~\ref{fig:bias1} we compare our results for the
bias function $b(L)$ with results by other authors.  Dashed line shows
our bias function (\ref{bias000}).  Black, blue and red lines show
fits by \citet{Norberg:2001aa}, \citet{Tegmark:2004aa} and
\citet{Zehavi:2011aa}.

\citet{Tegmark:2004aa} calculated power spectra in six bins of
absolute magnitude, and found that the relative bias function is
better given by the expression:
$f(L)=b(L)/b_\circ = 0.895+ 0.150(L/L_\ast) -0.040(M-M_\ast)$, see
their Figs.~28 and 29.  Here
$M_r=M_\ast -2.5\times \log_{10}(L/L_\ast)$ and $M_\ast$ are
$r-$magnitudes of SDSS galaxies and respective Schechter magnitudes.

\citet{Zehavi:2011aa} investigated the galaxy clustering of the
completed SDSS survey, and found for the  bias function the
form $b_g(>L) \times (\sigma_8/0.8) = 1.06 + 0.21(L/L_\ast)^{1.12}$,
where $L$ is the $r-$band luminosity corrected to $z=0.1$, and
$L_\ast$ corresponds to $M_\ast=-20.44 \pm 0.01$
\citep{Blanton:2003aa}.   \citet{Pollina:2018ul} found relative
linear bias factors between clusters and galaxies using first years of
observations of the Dark Energy Survey,
$b_{\mathrm{cl}}/b_{\mathrm{gal}} =1.6$ for $L>0.5L_\ast$ galaxies.

\citet{Shi:2016mz, Shi:2018bh} developed a method to map real space
distribution of galaxies.  The method was applied to measure
clustering amplitude of matter and bias parameters of flux-limited
sample of galaxies in SDSS DR7 in redshift range $0.01 \le z \le
0.2$. We show in
Fig.~\ref{fig:bias1} the bias parameter values, found by
\citet{Shi:2018bh} and applying normalising factor $b_\circ=1.4$.

\citet{Gil-Marin:2015yq, Gil-Marin:2017ly} investigated the clustering
of galaxies in the SDSS-III Baryon Oscillation Spectroscopic (BOSS)
Survey.  The BOSS survey selects LRG galaxies and consists of near and
distant samples, the LOWZ sample of effective redshifts
$z_{LOWZ} = 0.32$, and the distant sample CMASS with
$z_{CMASS} = 0.57$.  \citet{Gil-Marin:2017ly} found the linear biasing
parameter $b_1 = 2.08$ for LOWZ survey, and $b_1 = 2.01$ for CMASS
survey.  \citet{Gil-Marin:2017ly} compared observed bias parameters
with bias parameters for $N$-body halos of mock samples: low-bias
model with halo mass limit
$M_{min} = 3.80 \times 10^{12}~M_\odot h^{-1}$ has $b_1 = 1.75$, and
high-bias model with $M_{min} = 8.36 \times 10^{12}~M_\odot h^{-1}$
has $b_1 = 2.07$; see Table~1 by \citet{Gil-Marin:2017ly}.  These
values are close to bias parameter values found for our HR4.123 and
HR4.13 samples, see Table~\ref{Tab4} and Fig.~\ref{fig:Fig9}.  We show
in Fig.~\ref{fig:bias1} the bias parameter for BOSS LOWZ galaxies
according to \citet{Gil-Marin:2017ly}.  BOSS galaxies were selected
using LRG galaxies. We found for these galaxies the mean red magnitude
$M_r - 5\log h = -21.0$, with a spread about one magnitude.  Our SDSS
galaxy samples contain galaxies with luminosities greater or equal to
absolute magnitude limits. Thus we can accept for LRG galaxies the
magnitude limit  $M_r - 5\log h = -20.5$, almost equal to the
magnitude of $L_\ast$ galaxies. 
    
Power spectra of X-ray detected clusters of galaxies were derived in
the framework of REFLEX survey by \citet{Schuecker:2001ly}.  Power
spectra have a maximum around $k = 0.05~h$~Mpc$^{-1}$, shown in
Fig.~\ref{fig:bias} for two flux-limited cluster samples, L050 with
X-ray luminosity limit, $L_X \ge 0.5 \times 10^{44}$ erg s$^{-1}$, and
L120 with limit $L_X \ge 1.2 \times 10^{44}$ erg s$^{-1}$.  The
amplitude of the spectrum is higher for higher X-ray limit clusters.
Both spectra correspond to our model samples with very high particle
density limit $\delta_0 \approx 50$; see Fig.~\ref{fig:bias}. 
For X-ray clusters we estimated  the mean magnitude $M_r - 5\log h =
-21.5$.

\section{Discussion}

\begin{figure}  
\centering 
\hspace{2mm}  
\resizebox{0.42\textwidth}{!}{\includegraphics*{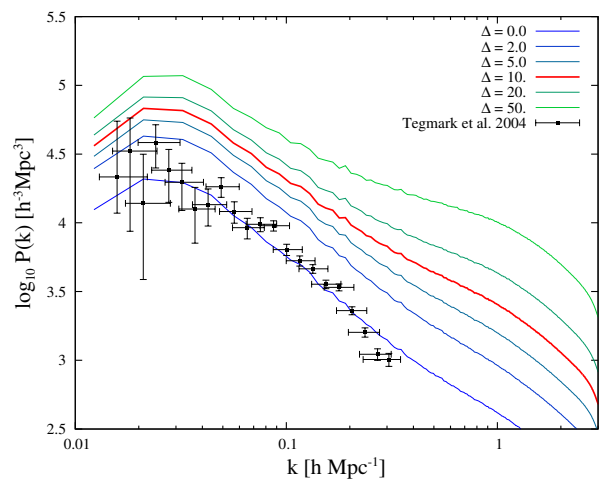}}
\hspace{2mm}  
\caption{Power spectra of particle density limited L512 model
  spectra.  The power spectrum of the sample L512.10 is shown by bold
  red line.  With filled symbols we show the power spectrum of SDSS
  galaxies by \citet{Tegmark:2004aa}. 
    }
\label{fig:bias10} 
\end{figure}

We show in Fig.~\ref{fig:bias10} power spectra of L512 models again;
the spectrum for biased model L512.10 is highlighted by red colour.
This biased model corresponds approximately to $L_\ast$ luminosity
limited galaxy samples.  We show in Fig.~\ref{fig:bias10} also the
power spectrum obtained by \citet{Tegmark:2004aa} according to their
Table~3, which corresponds to $L_\ast$ SDSS galaxies.

When the presence of the cosmic web was first discussed in the IAU
Tallinn Symposium, Zeldovich in his talk emphasised the importance to
develop statistical tools to measure quantitatively the new phenomena:
the filamentary character of the galaxy distribution and the presence
of voids.  So far the basic quantitative descriptor of the
distribution of galaxies was the two-point projected correlation
function $w_p(r_p)$.  This function was adequate to describe
two-dimensional data, as presented by \citet{Seldner:1977aa} and
analysed by \citet{Soneira:1978fk}.  Following this initiative
\citet{Zeldovich:1982kl} applied the percolation analysis to test the
filamentarity of the web, and the multiplicity analysis to test the
clustering properties.  The \citet{Soneira:1978fk} model failed in both
tests.  The Zeldovich own model, based on a HDM simulation by
\citet{Doroshkevich:1982fk}), failed in multiplicity test. Both tests
showed the agreement of model with observations only when a CDM model
was used \citep{Melott:1983}.

The cosmic web is very rich in details and has complex properties. For this reason
there exists no statistical tools which can describe all properties of
the web.  Each statistical tool is an instruments to test certain well
fixed properties of the web. To evaluate possible strengths and limits
of power spectrum determinations by various authors we have to
understand what features of the web can be tested by particular tools.
General properties of the cosmic web as delineated by galaxies and DM
are known long ago.  As discussed above, the main difference between
distributions of matter and galaxies is the presence of DM in
low-density regions, with no corresponding population of galaxies.

\begin{figure*}  
\centering 
\hspace{2mm}  
\resizebox{0.95\textwidth}{!}{\includegraphics*{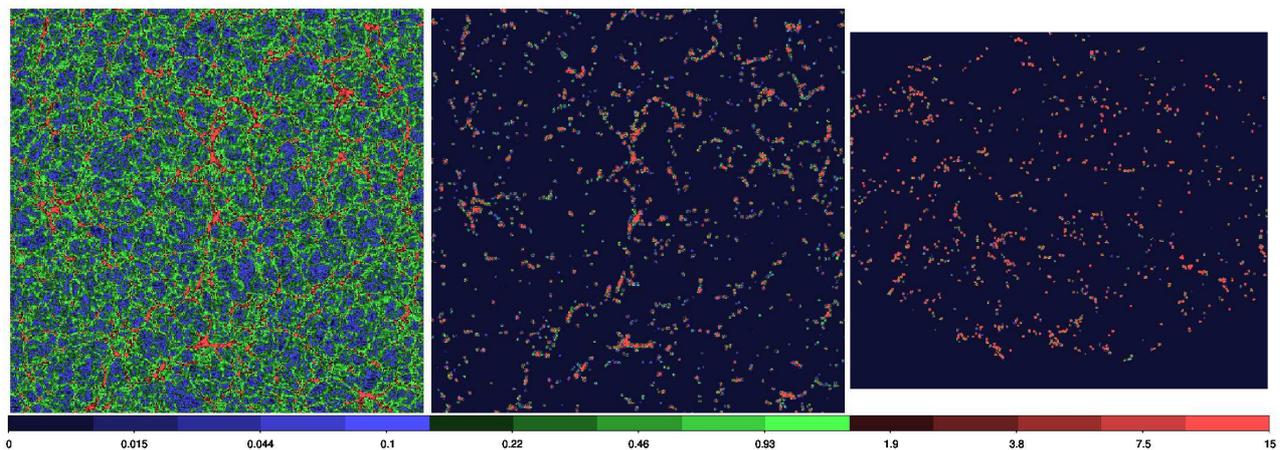}}\\
\caption{{\em Left  and central panels} show high-resolution
  non-smoothed density fields of identical
  $512\times 512 \times 1$~\Mpc\ slices  of dark
  matter models L512.00, L512.10, found for particle density limits
  $\delta_0=0,~10$. {\em Right panel} shows central section of the
  non-smoothed density field of the SDSS.21 galaxy sample. This Figure
  illustrates the effect of zero density regions in simulated and real
  density fields of galaxies.  Densities are expressed in
  logarithmic scale in interval 0.005 to 15 in mean density units.
  The colour code is identical in all panels.  }
\label{fig:Fig11} 
\end{figure*} 

\begin{figure*}
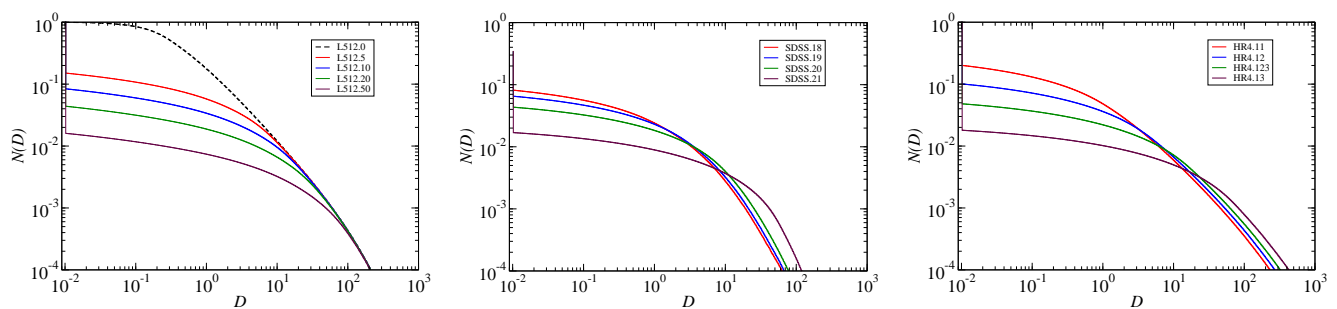
  
\centering 
\hspace{2mm}  
\resizebox{0.30\textwidth}{!}{\includegraphics*{L512_dens_int.eps}}
\hspace{2mm}
\resizebox{0.30\textwidth}{!}{\includegraphics*{SDSS_dens_int.eps}}
  \hspace{2mm}
\resizebox{0.30\textwidth}{!}{\includegraphics*{HR4_dens_int.eps}}\\
\hspace{2mm}
\caption{Cumulative distributions of densities in L512,  SDSS and HR4 
  samples, {\em left}, {\em middle} and {\em right panel} respectively.}
\label{fig:Fig13} 
\end{figure*} 

Geometrical properties of density fields of matter, model galaxies,
and SDSS galaxies were discussed in previous Sections. Here we discuss
some aspects of the distribution, critical to power spectrum analysis.
In Fig.~\ref{fig:Fig11} we compare high-resolution density fields of
these samples, represented by the full model sample L512.00, the
biased model sample L512.10, and the SDSS.21 sample.  $L_\star$
galaxies have approximately the magnitude $M_r - 5\log h = -20.5$,
thus the luminosity density field of SDSS.21 contains a bit higher
luminosity galaxies than the L512.10 field.  The Figure shows that
qualitatively the pattern of the distribution of simulated and real
clusters is similar.  The presence of large regions with zero spatial
density is well seen in simulated  and  real galaxy density
fields.

In Fig.~\ref{fig:Fig13} we compare cumulative distributions of
densities in density fields of model samples L512 with respective distributions for
observed SDSS samples and comparison HR4 samples.  For  the L512 model
 distributions are shown for the full sample with all particles,
L512.00, and for biased model samples L512.5, L512.10, L512.20 and
L512.50.  In the full sample there are particles with all density
labels, thus the cumulative distribution approaches unity with
decreasing particle density smoothly.  In all biased samples
low-density particles are absent, thus a large fraction of cells of
the density field have zero density.  The cumulative distribution has
a peak at lowest density value, corresponding to pixels with zero
density, and continues at lower level.  Density distributions of SDSS
and HR4 samples are qualitatively  similar to distributions of biased L512
samples.

\begin{figure}  
\centering 
\resizebox{0.45\textwidth}{!}{\includegraphics*{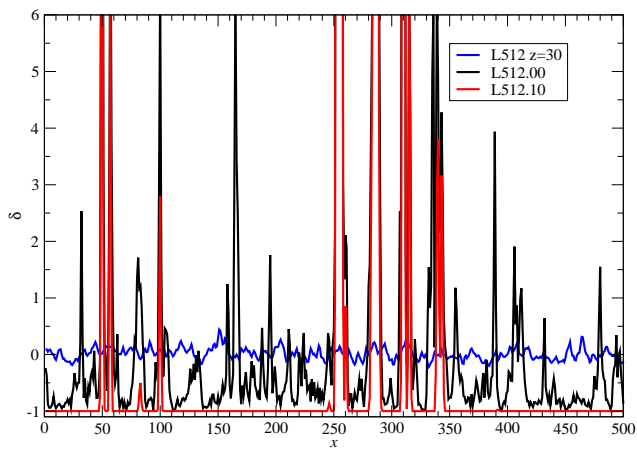}}
\caption{ 
 Cross sections of the density contrast $\delta(x) = D(x) - 1$
  of the model L512 along the $x$-spatial coordinate at the same
  $z$-coordinate as shown in Fig.~\ref{fig:Fig11}. Blue line shows the
  density field of the model L512.00 at redshifts $z=30$. Black line
  shows the density field of the full model L512.00 at the present
  epoch, and red line for the biased model L512.10.
}
\label{fig:Fig15} 
\end{figure}

Figure~\ref{fig:Fig15} shows a cross sections of the density field at
spatial $y,z$-coordinates through the center of the field, presented
in Fig.~\ref{fig:Fig11}.  We plot here the density contrast
$\delta(x) = D(x)-1$.  Blue line shows the density contrast at the
early epoch, corresponding to redshift $z=30$.  At this early epoch
the amplitude of density fluctuations is small, density fluctuations
are approximately equal everywhere.  Black line gives the density
contrast for the same cross section at the present epoch for the
sample with all particles, L512.00, and red line for the biased sample
L512.10.

In the sample L512.10 high-density peaks of the density field of
biased models are the same as in the full model.  Weak DM knots of
medium density, seen in the sample L512.00, are gone.  In low density
regions with $\delta < 6$ the density contrast of the L512.00 sample
fluctuates between values $0 < \delta < 6$ with a mean around
$\delta =-0.5$.  Over most of the density field the density of the
sample L512.10 has zero density and density contrast $\delta =-1$.

We give in Table~\ref{Tab2} the fraction of particles in biased
samples, $F_C$, and filling factor $FF_C$ of all clusters (non-zero
density cells) of the density field at threshold density $D_t=0.1$.
Both quantities are given as functions of the particle density limit
$\delta_0$ of biased model samples. The Table shows that the filling
factor of clusters decreases with increasing $\delta_0$ much faster
than the fraction of particles.  This is a well-known effect: the
density of clusters increases towards their centres and the volume
decreases more rapidly than the number of particles. For comparison we
note that filling factors of all clusters at the threshold density
level $D_t=0.1$ of samples SDSS.18,~SDSS.19, SDSS.20 and SDSS.21 are
0.1795,~0.1449,~0.0976 and 0.0399, respectively.  The fraction of zero
density cells is respectively increasingly closer to unity when we
increase the sample particle density limit $\delta_0$ or the
luminosity limit $L$.  In the sample L512.10 95\% of all cells of the
density field have zero density, in the SDSS.21 field even 96\% of
cells.

In our method power spectra are calculated using $\Lambda$CDM models.
Power spectrum is a sum over all density contrasts.  The sum is the
larger the greater is the fraction of cells with zero density and
density contrast $\delta = -1$.  For this reason the power spectrum
of all biased model samples has a higher amplitude than the full DM
model; the amplitude is the higher the larger the fraction of zero
density regions.  Thus the power spectrum is a measure of the fraction
of zero-density cells in the sample. It is interesting to note that
\citet{Einasto:1986oh} received the same conclusion from the analysis
of the three-dimensional correlation function.

A summary of measurements of power spectra is presented in
Figs.~\ref{fig:bias1} and \ref{fig:bias10}.  The power spectrum for
our full DM model L512.00 is in very good agreement with the updated
matter power spectrum at $z=0$, as compiled by
\citet{Chabanier:2019gb}.  The comparison of various determinations
shows that all methods permitted to determine very accurately the
luminosity dependence of galaxy power spectra, when appropriate bias
normalising factors $b_\circ$ are applied.  The amplitude of the
spectrum can be characterised by the bias normalising factor. Figures
show that normalising factors can be divided into two groups: around
$b_\circ \approx 1$ --- \citet{Norberg:2001aa}, \citet{Tegmark:2004aa} and
\citet{Zehavi:2011aa}; and around $b_\circ \approx 2$ ---
\citet{Schuecker:2001ly}, \citet{Gil-Marin:2015yq, Gil-Marin:2017ly}
and our work.  This variety of bias normalising factors suggests that
authors used different tools to handle zero density regions of the
luminosity density field of galaxies.

Devil is in detail.  Methods to estimate power spectra of galaxies
contain numerous technical details and assumptions.  Each method yields
results what the method permits. In this respect a combined method
using different properties of the cosmic web has an advantage to see
the bias phenomenon in a broader context.  It is clear that future
development adds new details to the picture we have today.

\section{Concluding remarks}

The present study shows that the absence of galaxy formation in
low-density regions of the cosmic web is an essential property of the
$\Lambda$CDM universe. It follows from the combined action of several
physical processes: (i) the smoothness of the flow of particles until
the intersection of particle trajectories; (ii) the formation of halos
along caustics of particle trajectories; (iii) the phase
synchronisation of density perturbations of various scales; and
(iv) the two-step nature of galaxy formation by condensation of
baryonic matter within DM halos.

We studied the distribution of galaxies and matter using
respective density fields and applying percolation and power spectrum
analyses.  To calculate the density fields of simulated galaxies we
used the particle local density $\delta$ as a dimensionless
characteristic of particles of numerical simulations of the cosmic
web.  We applied sharp particle density limit, $\delta \ge \delta_0$,
to select particles, which form biased model samples.  We tested this
selection algorithm using fuzzy particle density limits and analysing
number functions of simulated and real galaxies.  Our analysis
shows that this sample selection method yields biased model samples in
a wide range of particle density limits, and allows to calculate bias
function of biased model samples as functions of the particle density
limit, $b(>\delta_0)$. The bias function depends on cosmological
parameters of the model only weakly, since we use ratios of power
spectra of the same model.

We compared biased model samples with luminosity selected SDSS galaxy
samples using the extended percolation analysis. Our analysis shows
that the extended percolation method allows to compare observational
and model samples having very different sample sizes and
configurations.  The method is almost not influenced by redshift space
distortions, present in observational samples. The percolation method
is very sensitive to geometrical properties of clusters and voids of
observed and model samples, and allows to find density limits
$\delta_0$ of biased models, which correspond to luminosity limited
SDSS samples..

As a result of physical processes described above there exists at all
cosmological epochs a low-density population, consisting of a mixture
of dark and baryonic matter.  A crucial role in the evolution of the
universe plays the phase synchronisation which leads to the formation
of small filamentary high-density regions and large contiguous regions
with very low spatial densities.  Galaxy formation is possible only in
the high-density medium.  This is the main factor in the biasing
phenomenon, leading to increase of zero density cells in density field
of simulated and real galaxies, and an increase of the amplitude of
the power spectrum of galaxies in respect to the power spectrum of
matter.  The second largest effect is the dependence of the bias
function on the luminosity of galaxies.  Variations in gravitational
and physical processes during the formation and evolution of galaxies,
represented in our biasing model by the fuzziness of the biasing
threshold, have the smallest influence.

The combined geometrical and power spectrum analysis demonstrates well
the presence of large differences between distributions of matter and
galaxies, expressed quantitatively by percolation functions and power
spectra. Power spectra of biased models representing SDSS samples of various
luminosity limits allowed to calculate the expected bias function, $b(>L)$.  
The bias function is in very good agreement with earlier studies when
appropriate bias normalising factors are applied.

Our main conclusions are.

\begin{enumerate}

\item  Non-clustered matter in low-density regions is smoothly
    distributed, which rises the amplitude of power spectra of the
    clustered matter in galaxies in respect to the amplitude of power
    spectra of all matter.  This is the dominant factor to influence
    the biasing phenomenon, and can be used as a cosmological constraint.
  
\item  The dependence of the bias parameter on the luminosity of
    galaxies is the second largest effect affecting the bias
    parameter.  Variations in gravitational and physical processes
    during the formation and evolution of galaxies have the smallest
    influence to the bias parameter.
  
\item  Combined analysis of geometrical properties of the cosmic web
    and power spectra of biased model samples and SDSS samples of
    galaxies allow to estimate the bias parameter of $L_\ast$ galaxies,
    $b_\ast =1.85 \pm 0.15$.

\end{enumerate}

\begin{acknowledgements} 

  Our special thank is to Gert H\"utsi for calculations of power
  spectra and many stimulating discussions.
  We thank Changbom Park for
  the permission to use Horizon 4 simulations for this study, and Mirt
  Gramann, Enn Saar, Antti Tamm, Elmo Tempel and Rien van de Weygaert
  for discussion \footnote{We tell with sorrow that during  final preparations of
  the paper  our collaborator Ivan Suhhonenko
  (1974 -- 2019) passed away.}.

This work was supported by institutional research funding IUT26-2 and
IUT40-2 of the Estonian Ministry of Education and Research. We
acknowledge the support by the Centre of Excellence ``Dark side of the
Universe'' (TK133) financed by the European Union through the European
Regional Development Fund.  The study has also been supported by
ICRAnet through a professorship for Jaan Einasto.

We thank the SDSS Team for the publicly available data releases.
Funding for the SDSS and SDSS-II has been provided by the Alfred
P. Sloan Foundation, the Participating Institutions, the National
Science Foundation, the U.S. Department of Energy, the National
Aeronautics and Space Administration, the Japanese Monbukagakusho, the
Max Planck Society, and the Higher Education Funding Council for
England. The SDSS Web Site is \texttt{http://www.sdss.org/}.

The SDSS is managed by the Astrophysical Research Consortium for the 
Participating Institutions. The Participating Institutions are the 
American Museum of Natural History, Astrophysical Institute Potsdam, 
University of Basel, University of Cambridge, Case Western Reserve 
University, University of Chicago, Drexel University, Fermilab, the 
Institute for Advanced Study, the Japan Participation Group, Johns 
Hopkins University, the Joint Institute for Nuclear Astrophysics, the 
Kavli Institute for Particle Astrophysics and Cosmology, the Korean 
Scientist Group, the Chinese Academy of Sciences (LAMOST), Los Alamos 
National Laboratory, the Max-Planck-Institute for Astronomy (MPIA), 
the Max-Planck-Institute for Astrophysics (MPA), New Mexico State 
University, Ohio State University, University of Pittsburgh, 
University of Portsmouth, Princeton University, the United States 
Naval Observatory, and the University of Washington.

\end{acknowledgements}

\bibliographystyle{aa} 

\end{document}